# Statistical analysis of hard X-ray radiation at PAL-XFEL facility performed by Hanbury Brown and Twiss interferometry

Authors

**Young Yong Kim[a], Ruslan Khubbutdinov[a], Jerome Carnis[a], Sangsoo Kim[b], Daewoong Nam[bc], Inhyuk Nam[b], Gyujin Kim[b], Chi Hyun Shim[b], Haeryong Yang[b], Myunghoon Cho[b], Chang-Ki Min[b], Changbum Kim[b], Heung-Sik Kang[b] and Ivan Vartanyants[a]***

[a]Photon Science, Deutsche Elektronen-Synchrotron DESY, Notkestr. 85, Hamburg, 22607, Germany

[b] Pohang Accelerator Laboratory, Pohang, Gyeongbuk, 37673, Republic of Korea

[c]Photon Science Center, POSTECH, Pohang, 37673, Republic of Korea

Correspondence email: Ivan.Vartaniants@desy.de

**Funding information**    Ministry of Science and Information, Communications & Technology (ICT) of Korea (grant No. 2018R1A6B4023605); Basic Science Research Program (grant No. NRF-2021R1A2C3010748 ; grant No. 2020R1C1C1011840); National Research Foundation of Korea (grant No. NRF-2019R1A6B2A02098631 to D.N.; grant No. NRF-2021R1F1A1051444 to D.N.).

**Synopsis**    Statistical properties of hard X-ray free-electron laser PAL-XFEL were studied by Hanbury Brown and Twiss interferometry. Our results demonstrate high spatial coherence and short average pulse duration of this facility at 10 keV photon energy.

**Abstract**    Hanbury Brown and Twiss interferometry experiment based on second-order correlations was performed at PAL-XFEL facility. The statistical properties of the X-ray radiation were studied within this experiment. Measurements were performed at NCI beamline at 10 keV photon energy in various operation conditions: Self-Amplified Spontaneous Emission (SASE), SASE with a monochromator, and self-seeding regimes at 120 pC, 180 pC, and 200 pC electron bunch charge, respectively. Statistical analysis showed short average pulse duration from 6 fs to 9 fs depending on operation conditions. A high spatial degree of coherence of about 70-80% was determined in spatial domain for the SASE beams with the monochromator and self-seeding regime of operation. The obtained values describe the statistical properties of the beams generated at PAL-XFEL facility.

## 1. Introduction

Hard X-ray free-electron lasers (XFEL) are currently the brightest X-ray sources in the world (Emma *et al.*, 2010; Ishikawa *et al.*, 2012; Kang *et al.*, 2017; Decking *et al.*, 2020; Prat *et al.*, 2020). These facilities provide intense hard X-ray beams with high coherence properties and pulse duration in the range of tens to hundred femtoseconds (Mcneil & Thompson, 2010; Pellegrini *et al.*, 2016). Such unique properties have triggered research in atomic physics (Young *et al.*, 2010; Rohringer *et al.*, 2012; Prince *et al.*, 2016), structural dynamics that determine the function of proteins (KERN *et al.*, 2013; Nogly *et al.*, 2018), mechanisms controlling chemical bonds during catalytic reactions (Dell'Angela *et al.*, 2013; Öström *et al.*, 2015), and the processes that are interesting for energy conversion and information storage applications (Beaud *et al.*, 2014; Dornes *et al.*, 2019). The high peak intensities and short pulse duration generated by these facilities have introduced entirely new fields of research such as femtosecond crystallography (Chapman *et al.*, 2011) and single particle imaging (SPI) (Seibert *et al.*, 2011; Aquila *et al.*, 2015), allowing the determination of a three-dimensional biological particle with a resolution of less than 10 nm (Rose *et al.*, 2018; Assalauova *et al.*, 2020). Furthermore, the advantages of the XFEL have allowed to perform such coincidence-based experiments as incoherent and ghost imaging (Schneider *et al.*, 2018; Kim *et al.*, 2020).

A crucial factor in generating the unique characteristics of XFEL radiation is the X-ray lasing process. The key principle utilized at most of the XFEL facilities is based on the self-amplified spontaneous emission (SASE) process, allowing the generation of high intense XFEL pulses (Saldin *et al.*, 2000; Milton *et al.*, 2001). The beams generated by SASE radiation have high degree of spatial coherence and many longitudinal modes that vary randomly from one pulse to another, and statistically such XFELs behave as a chaotic source (Singer *et al.*, 2013; Gorobtsov *et al.*, 2017a; Gorobtsov, Mukharamova, Lazarev, Chollet, Zhu, Feng, Kurta, Meijer, Williams, Sikorski, Song *et al.*, 2018; Khubbutdinov *et al.*, 2021). An important exception from this rule is the externally seeded FEL at FERMI facility in Trieste (Italy), which behaves as a truly one-mode laser source with high degree of spatial coherence (Gorobtsov, Mercurio *et al.*, 2018).

As it was demonstrated at the Linac Coherent Light Source (LCLS), one of the ways to obtain a narrow spectral line about the size of a single spike in the SASE spectrum is to perform self-seeding (Amann *et al.*, 2012). The self-seeding at hard XFELs is based on installing a diamond crystal of high quality in Bragg geometry instead of the undulator section in the undulator line (Geloni *et al.*, 2011). After successful commissioning of hard X-ray radiation generated by SASE at Pohang Accelerator Laboratory X-ray Free-Electron Laser (PAL-XFEL) in 2016 (Kang *et al.*, 2017), the self-seeding operation was also successfully implemented in 2018 (Min *et al.*, 2019; Nam *et al.*, 2021). Now, a natural question is: what are the statistical properties of the self-seeded X-ray beams from the XFEL sources? Are they laser-like as in the case of externally seeded FELs (Allaria *et al.*, 2013) or do they have rather chaotic nature as SASE FELs?

In order to answer this fundamental question about the statistical properties of self-seeded XFELs, one may use the method of Hanbury Brown and Twiss (HBT) interferometry. The method is based on the second-order intensity correlations and was first introduced experimentally by Hanbury Brown and Twiss (Hanbury Brown & Twiss, 1956; Brown & Twiss, 1956). Later, it led to creation and development of the field of quantum optics (Glauber, 1963; Sudarshan, 1963). Currently, this method has been successfully applied for the analysis of X-ray radiation at different FEL facilities (Singer *et al.*, 2013; Gorobtsov *et al.*, 2017a; Gorobtsov, Mukharamova, Lazarev, Chollet, Zhu, Feng, Kurta, Meijer, Williams, Sikorski, Song *et al.*, 2018; Inoue *et al.*, 2018; Khubbutdinov *et al.*, 2021).

In this work, we present a statistical analysis of the hard X-ray beams generated by PAL-XFEL in different operation conditions using HBT interferometry. These conditions are: SASE radiation, SASE radiation with the monochromator, and self-seeding regime of operation. The latter is of particular interest in terms of understanding the self-seeding operational mode of this facility.

## 2. HBT interferometry

The HBT interferometry is a method that uses the second-order correlation of intensity measured in spatial or temporal domains and is effective in analysing the statistical properties of the optical wave fields. The normalized second-order correlation function in spatial domain is expressed as

$$g^{(2)}(\boldsymbol{r_1}, \boldsymbol{r_2}) = \frac{\langle I(\boldsymbol{r_1})I(\boldsymbol{r_2})\rangle}{\langle I(\boldsymbol{r_1})\rangle\langle I(\boldsymbol{r_2})\rangle}, \qquad (1)$$

where $I(\boldsymbol{r_1})$, $I(\boldsymbol{r_2})$ are the intensities of the wave field in spatial domain and averaging denoted by brackets $<\ldots>$ is performed over a large ensemble of different realizations of the wave field. Similar expression will hold in spectral domain.

If radiation is cross-spectrally pure and obeys Gaussian statistics, which means it is analogous to a chaotic source (Mandel & Wolf, 1995), the $g^{(2)}$ - function may be expressed as (Ikonen, 1992; Singer *et al.*, 2013; Vartanyants & Khubbutdinov, 2021),

$$g^{(2)}(\boldsymbol{r_1}, \boldsymbol{r_2}) = 1 + \zeta_2(D_\omega) \cdot \left| g^{(1)}(\boldsymbol{r_1}, \boldsymbol{r_2}) \right|^2, \qquad (2)$$

where $g^{(1)}(\boldsymbol{r_1}, \boldsymbol{r_2}) = \langle E^*(\boldsymbol{r_1})E(\boldsymbol{r_2})\rangle / \sqrt{I(\boldsymbol{r_1})I(\boldsymbol{r_2})}$ is the first-order correlation function or spectral degree of coherence and $\zeta_2(D_\omega)$ is the contrast function, which depends on radiation bandwidth $D_\omega$. The contrast, $\zeta_2(D_\omega)$, is proportional to $\tau_c/T$ in the limit when the average pulse duration ($T$) is much larger than the coherence time ($\tau_c$) ($T \gg \tau_c$). Conversely, if the coherence time is larger than the pulse duration, the contrast has a constant value close to one (Singer *et al.*, 2013; Vartanyants & Khubbutdinov, 2021).

## 3. Results

### 3.1. Experiment

The HBT experiment was performed at the Nano-crystallography and coherent imaging (NCI) hard X-ray beamline at Pohang Accelerator Laboratory X-ray Free-Electron Laser (PAL-XFEL) (Park *et al.*, 2016; Kang *et al.*, 2017). The PAL-XFEL was operated at 10 GeV electron energy with the three different electron bunch charges of 120 pC, 180 pC, and 200 pC with 30 Hz repetition rate. The schematic image of the experimental set-up is shown in Fig. 1. The X-ray photon energy for the experiment was 10 keV ($\lambda$ = 1.24 Å) with 20 sections of undulators which were 5 m in length in saturation regime (Ko *et al.*, 2017). For all bunch charges the experiment was performed with SASE radiation, SASE radiation with the monochromator, and self-seeded radiation modes. In addition to these modes, linear regime was used in few cases with 12 undulator sections for 120 pC bunch charge and 13 undulator sections for 200 pC bunch charge. Typical recorded data of SASE single pulses are shown in Fig. 2 for the 180 pC bunch charge and in the Supplementary Information (SI) Fig. S1 for the bunch charge 120 pC.

For the monochromatic operation, a double-crystal Si (111) monochromator (DCM) was installed, which was positioned 99.84 m downstream from the source point. The theoretical resolution of the DCM was $\Delta E/E$=1.865·10⁻⁴ at 10 keV photon energy (X-ray server). During the analysis of our experiment we observed vertical position drifts of the monochromator that were corrected by further analysis (see SI Fig. S2).

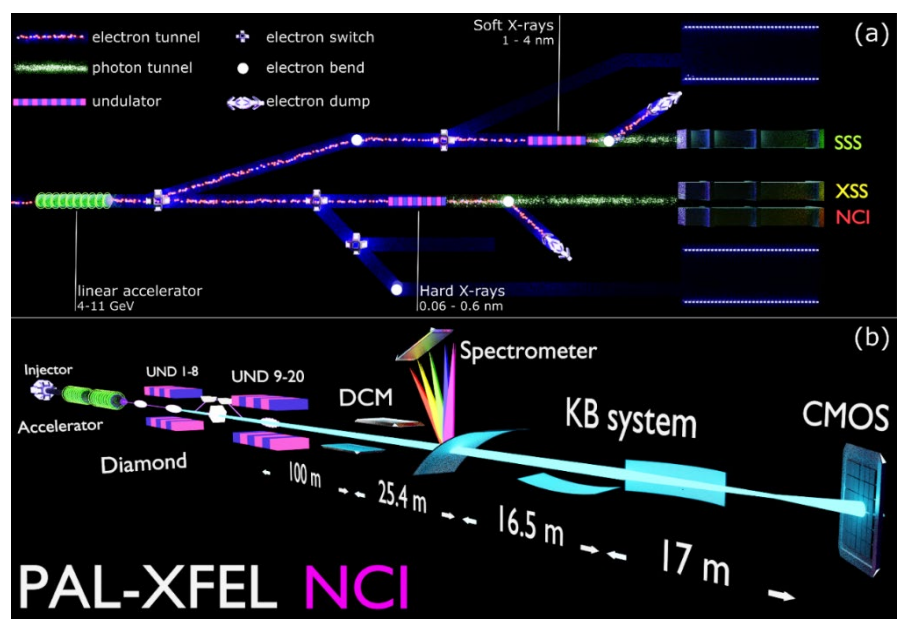

**Figure 1** Schematic image of the experimental set-up. (a) The general outline of the PAL-XFEL facility. (b) An outline of the Nano-crystallography and coherent imaging (NCI) beamline at PAL-XFEL. For the SASE radiation 20 undulator sections were used. In the self-seeding mode of operation, the diamond crystal between the 8-th and 9-th sections of undulator was implemented.

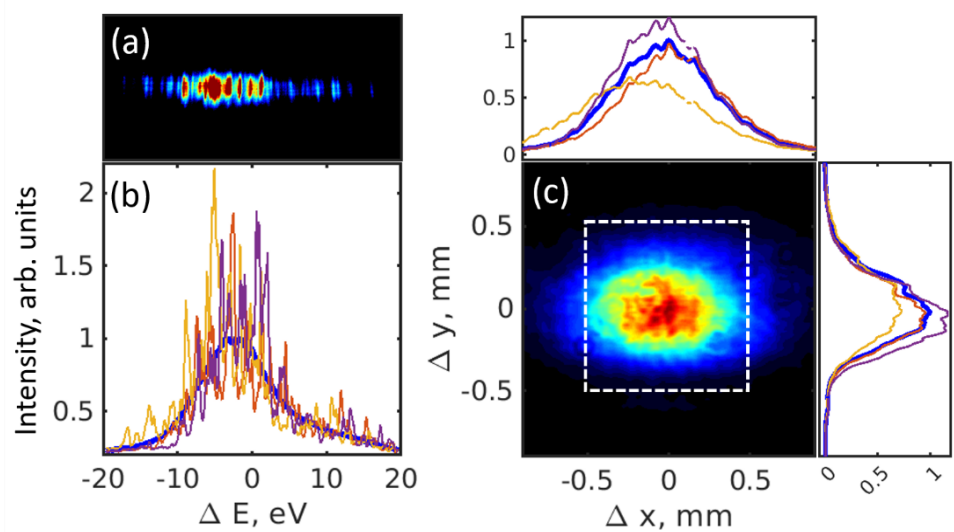

**Figure 2** Spectral and spatial intensity distributions for the SASE operating conditions with the 180 pC bunch charge. (a) Typical 2D spectral intensity distribution of a single pulse in the SASE mode measured by the Andor CCD detector after subtraction of the dark images. (b) Projection of several individual pulses in the vertical direction and an average spectrum (blue line). (c) An average spatial intensity distribution in the SASE mode measured by the Hamamatsu detector. The region of interest for the correlation analysis in spatial domain is shown in (c) by the white dashed square. Projection of individual pulses as well as an average intensity distribution (blue curve) in the horizontal and vertical directions is also shown in (c).

For the self-seeded operation, the forward Bragg diffraction (FBD) diamond monochromator was used, which was located after eight undulators and amplified with twelve undulators downstream (Min *et al.*, 2019; Nam *et al.*, 2021).

The spectrum of each pulse was measured by an on-line spectrometer. The spectrometer consists of a Si (333) bent crystal and an Andor detector (ZYLA5.5X-FO, 2560 × 400 pixels, pixels size 6.5 × 6.5 μm$^2$) positioned at 1.17 m from the bent silicon crystal (Ko *et al.*, 2017). The dispersion value at the position of the spectrometer detector was estimated to be 6 eV/mm. Resolution of the on-line spectrometer was estimated to be 0.26 eV (FWHM) (Nam *et al.*, 2021). The on-line spectrometer was located 25.4 m downstream from the DCM.

All spatial measurements were performed with the focused beam using Kirkpatrick-Baez (KB) mirrors located 5.37 m upstream from the focal position. The spatial beam intensities were measured by the Hamamatsu X-ray sCMOS camera (2048 × 2048 pixels, pixels size 6.5 × 6.5 μm$^2$). The region of interest, where data were collected during the experiment, was defined as 600 × 600 pixels. This detector was positioned 11.5 m downstream from the focal position. To prevent beam damage of the spectral and spatial detectors, a 0.28 mm thick silicon attenuator was positioned in front of the Andor detector and silicon attenuators of different thicknesses from 1.175 mm to 1.5 mm, depending on beam conditions, were positioned in front of the Hamamatsu detector.

## 3.2. Spectral analysis

Single-pulse spectra and intensities in spatial domain were collected simultaneously. To get statistically relevant results we collected from 8,000 to 20,000 pulses (see SI Table S1) at each operating condition of PAL XFEL. Each spectrum and intensity in spatial domain were corrected by the mean dark image with 1,000 shots. The one-dimensional single pulse spectrum was obtained by projection of the two-dimensional spectrum image along the vertical direction (see Fig. 2 and SI Fig. S1). By that, we obtained the single-pulse spectral intensity distribution, as well as an average spectrum for all operating conditions (see Fig. 3(a,c,e) for 180 pC bunch charge and for other bunch charges SI Figs. S3 and S4).

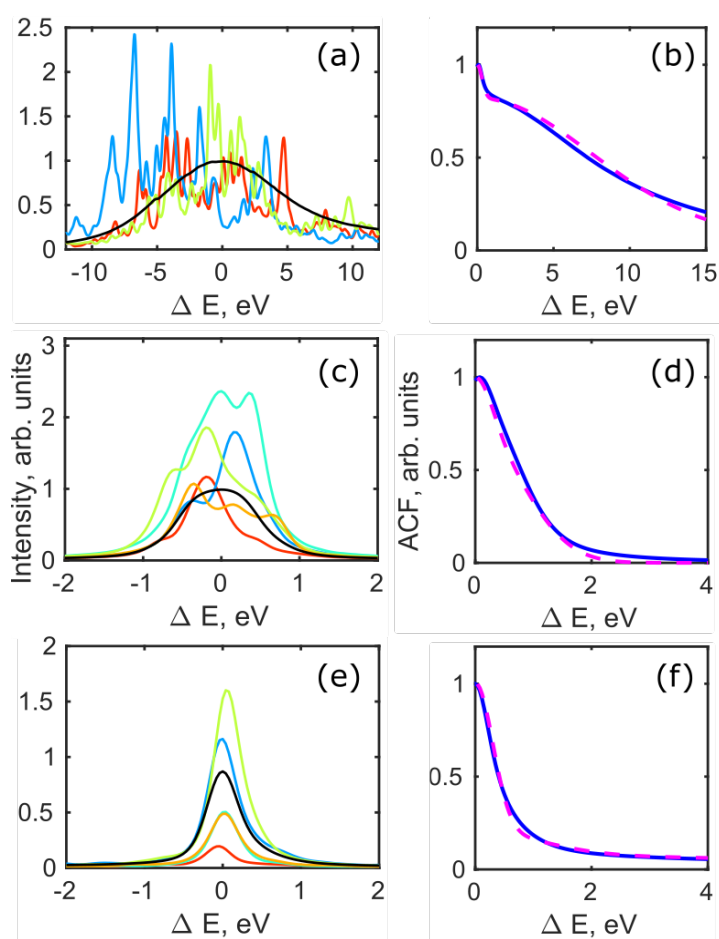

**Figure 3**  (a, c, e) Spectral distribution of few random pulses and an average spectrum for all pulses (black lines). (b, d, f) Average autocorrelation function of all spectral lines (blue lines), and its fit (red dashed lines) by two Gaussian functions. (a,b) SASE radiation, (c,d) monochromatic radiation, (e,f) self-seeding regime of operation. All results presented in this figure correspond to the 180 pC bunch charge.

From the average spectrum we estimated the full width at half maximum (FWHM) of the spectrum for all operation conditions (see Table 1). We observed that the width of the SASE spectrum was about 12 eV for 120 pC and 180 pC bunch charges and is close to a single Gaussian (this was similar to SASE operation in Refs. (Min *et al.*, 2019; Nam *et al.*, 2021)). Contrary to that the averaged SASE spectrum for 200 pC bunch charge was more than twice wider (~28 eV) and may be well represented by a sum of two Gaussian functions shifted in energy. Such different spectral behaviour at different bunch charges strongly depends on specific machine tuning by the XFEL operators. At the same time, for all three operation conditions, the monochromatic radiation has the same bandwidth of about 1.1 - 1.2 eV. This value is slightly narrower than provided by the theoretical bandwidth of the DCM at 10 keV ($\Delta E$=1.9 eV). The reason for this may be a slight detuning of two Si crystals from which DCM is composed. For the self-seeding regime of operation, we also observed the same behaviour for all three bunch charges, the average spectrum was extremely narrow and was about 0.4 eV (see Table 1).

**Table 1**   Results of the analysis in spectral domain for all bunch charges used in the experiment. XFEL spectrum bandwidth was obtained directly from the averaged spectrum of all pulses in each operation condition. Analysis of the autocorrelation function (ACF) was performed by Eq. (6) that provided the intrinsic XFEL spectrum bandwidth as well as the spike bandwidth. Coherence times were estimated from Eq. (5) assuming that the average spectrum is fitted by two Gaussian functions. Pulse duration was obtained by fitting the central part of the second-order correlation function $g_{in}(\Delta\omega)$ in Eq. (9). The values of different parameters measured in the linear mode of operation are also provided in the Table.

| Bunch charge | 120 pC | | | 180 pC | | | 200 pC | | |
|---|---|---|---|---|---|---|---|---|---|
| Operation mode | SASE radiation | Mono-chromatic radiation | Self-seeding radiation | SASE radiation | Mono-chromatic radiation | Self-seeding radiation | SASE radiation | Mono-chromatic radiation | Self-seeding radiation |
| XFEL spectrum bandwidth (FWHM), eV | 11.9 /13.3 (L) | 1.2  /1.2 (L) | 0.4 | 11.5 | 1.1 | 0.4 | 27.8 | 1.1 | 0.4 /0.5 (L) |
| XFEL spectrum bandwidth from ACF (FWHM), eV | 12.41±0.3/ 12.0±0.1 (L) | 1.37±0.1 /1.37±0.1 (L) | 3.09±0.1 | 12.6±0.1 | 1.3±0.1 | 2.9±0.1 | 24.0±0.2 | 1.3±0.2 | 3.6±0.1 /4.1±0.1 (L) |
| Spike bandwidth from ACF (FWHM), eV | 0.4±0.1 /0.4±0.1 (L) | 0.4±0.1 /0.4±0.1 (L) | 0.4±0.1 | 0.4±0.1 | 0.4±0.1 | 0.4±0.1 | 0.4±0.1 | 0.4±0.1 | 0.4±0.1 /0.54±0.1 (L) |
| Coherence time (rms), fs | 0.17±0.01 /0.17±0.01 (L) | 2.55±0.03 /2.56±0.03 (L) | 4.64±0.03 | 0.17±0.01 | 2.48±0.28 | 4.19±0.10 | 0.11±0.01 | 2.25±0.06 | 3.68±0.15 /3.44±0.04 (L) |
| Pulse duration T, fs | 6.0±0.2 /6.1±0.2 (L) | 7.2±0.2 /6.0±0.2 (L) | --- | 7.0±0.2 | 8.8±0.2 | --- | 6.4±0.2 | 7.2±0.2 | --- |

From the average spectrum we can determine the coherence time of the PAL XFEL radiation at different operation conditions. The coherence time is given by the following expression (Goodman, 2000; Mandel & Wolf, 1995; Khubbutdinov *et al.*, 2021)

$$\tau_c = \int_{-\infty}^{\infty} |\gamma(\tau)|^2 d\tau \ , \tag{3}$$

where $\gamma(\tau)$ is the complex degree of coherence. The complex degree of coherence may be determined through the average spectrum S(ω) as

$$\gamma(\tau) = \frac{\int_0^{\infty} S(\omega) e^{-i\pi\tau} d\omega}{\int_0^{\infty} S(\omega) d\omega}. \tag{4}$$

For simple spectral shapes coherence time is determined in Ref. (Goodman, 2000). If the spectral shape is given by a Gaussian function, then $\tau_c = \sqrt{\pi}/\sigma_\omega$, where $\sigma_\omega$ is the root mean square (rms) value of the Gaussian spectrum. Since the average spectrum in our experiment does not follow the shape of a single Gaussian function, we used the sum of two Gaussian functions to get an estimate of the coherence time of the radiation in all operating conditions (see SI Table S2 for parameters of these Gaussian functions). Coherence time for a spectrum modelled by a sum of two Gaussian functions may be expressed as (Khubbutdinov *et al.*, 2021)

$$\tau_c = \frac{\sqrt{\pi}}{[A_1\sigma_1 + A_2\sigma_2]^2} \left\{ A_1^2\sigma_1 + A_2^2\sigma_2 + \frac{2\sqrt{2}A_1A_2\sigma_1\sigma_2}{\sqrt{\sigma_1^2 + \sigma_2^2}} exp\left[ -\frac{\Delta\omega_0^2}{2(\sigma_1^2 + \sigma_2^2)} \right] \right\}, \tag{5}$$

where $A_1$ and $A_2$ are scale factors, $\Delta\omega_0 = \omega_2^0 - \omega_1^0$, $\omega_1^0$ and $\omega_2^0$ are the centres of each Gaussian line, and $\sigma_1$ and $\sigma_2$ are their rms values. The results of fitting the average spectrum by two Gaussian functions for all operation conditions are shown in SI Fig. S5. The determined values of the coherence time according to Eq. (5) are summarized in Table 1.

The values of coherence time were about 170 as for 120 pC and 180 pC bunch charges which are typical for hard X-ray SASE operation at different XFEL facilities (Vartanyants *et al.*, 2011; Gutt *et al.*, 2012; Lehmkühler *et al.*, 2014). For monochromatic radiation, coherence times increased up to 2.5 fs for both bunch charges. Due to a broader spectrum in the case of 200 pC bunch charge, the coherence times of SASE radiation were about twice shorter and about 110 as. Interestingly, for monochromatic radiation with the 200 pC bunch charge, coherence times were 2.2 fs, that are similar to other bunch charges. The latter is explained by the fact that the bandwidth of monochromatic radiation is given by the DCM, which was the same for all three bunch charges. In the case of self-seeding operation mode, the sharp spectrum was staying on the broad pedestal. This pedestal originates from longitudinal phase space modulations produced by the microbunching instability upstream of the undulators as well as SASE background (Nam *et al.*, 2021). In our estimates of the coherence time we used this sharp peak above the broad background which gave us, finally, about twice longer coherence times (~4 fs) in comparison to monochromatic radiation for all bunch charges.

The analysis of the averaged auto-correlation function (ACF) allowed us to determine the bandwidth of a single spike in single pulse spectra (Khubbutdinov *et al.*, 2021). The ACF in Fig. 3 were fitted by a sum of two Gaussian functions as

$$ACF(\Delta E) = \frac{1}{N} \sum_{i=1}^{N} S_i(\Delta E) \otimes S_i(\Delta E) = A_1 exp\left[-\frac{(\Delta E)^2}{4(\sigma_{spectr}^2 + \sigma_r^2)}\right]$$
$$+ A_2 exp\left[-\frac{(\Delta E)^2}{4(\sigma_{spike}^2 + \sigma_r^2)}\right], \tag{6}$$

where $S_i(\Delta E)$ is the individual spectral line measured by the on-line spectrometer for each pulse, $N$ is the number of pulses, $\otimes$ is a correlation sign, $A_1$ and $A_2$ are normalization constants, $\sigma_{spectr}$ is the rms value of an averaged spectrum, $\sigma_{spike}$ is the rms value of an average spike, and $\sigma_r$ is the rms value of the resolution of on-line spectrometer that was considered to be $\sigma_r$=0.11 eV (Nam *et al.*, 2021). We performed the ACF analysis for all three bunch charges and operation conditions studied at PAL XFEL and observed similar profiles for each operation mode regardless of the bunch charge (see Fig. 3 for the 180 pC bunch charge and for the other bunch charges SI Figs. S3 and S4). For SASE radiation for all bunch charges we clearly observed a sharp peak corresponding to the spike shape staying on the pedestal of a broad peak corresponding to the spectrum bandwidth (see Fig. 3(b) and SI Figs. S3 (b,d) and S4(b)). For the monochromatic beams we did not resolved individual spikes in the ACF and in the case of self-seeding a sharp peak corresponding to self-seeding radiation was staying on the broad pedestal. To determine the width of the peaks from the ACF's we performed the Gaussian fit of the ACF's according to Eq. (6) (see Table 1).

From our ACF analysis we obtained the width of the spike in the case of SASE, monochromatic, and self-seeding radiation to be about 0.4±0.1 eV (FWHM) for all three bunch charges used in our experiment at the PAL XFEL facility. We checked the obtained value by analysing the width of few individual pulses and obtained the same value.

We further analysed the second-order correlation function in the spectral domain

$$g^{(2)}(\omega_1, \omega_2) = \frac{<I(\omega_1 - \omega_0)I(\omega_2 - \omega_0)>}{<I(\omega_1 - \omega_0)><I(\omega_2 - \omega_0)>},$$ (7)

where $\omega_0$ is the central frequency.

The second-order correlation functions in the frequency domain for different operation conditions and 180 pC bunch charge are shown in Fig. 4 (see for the other bunch charges SI Figs. S9 – S12). As we clearly see from these results the behaviour of the $g^{(2)}(\omega_1, \omega_2)$ −function is similar for the same modes of operation disregarding the bunch charge. For SASE regime of operation, it has a narrow peak along the main diagonal with two maxima in the bottom left and top right positions (see Fig. 4(a) and SI Figs S9(a) and S10(a)). These maxima are indication of the energy jitter as described in Ref. (Gorobtsov *et al.*, 2017*b*).

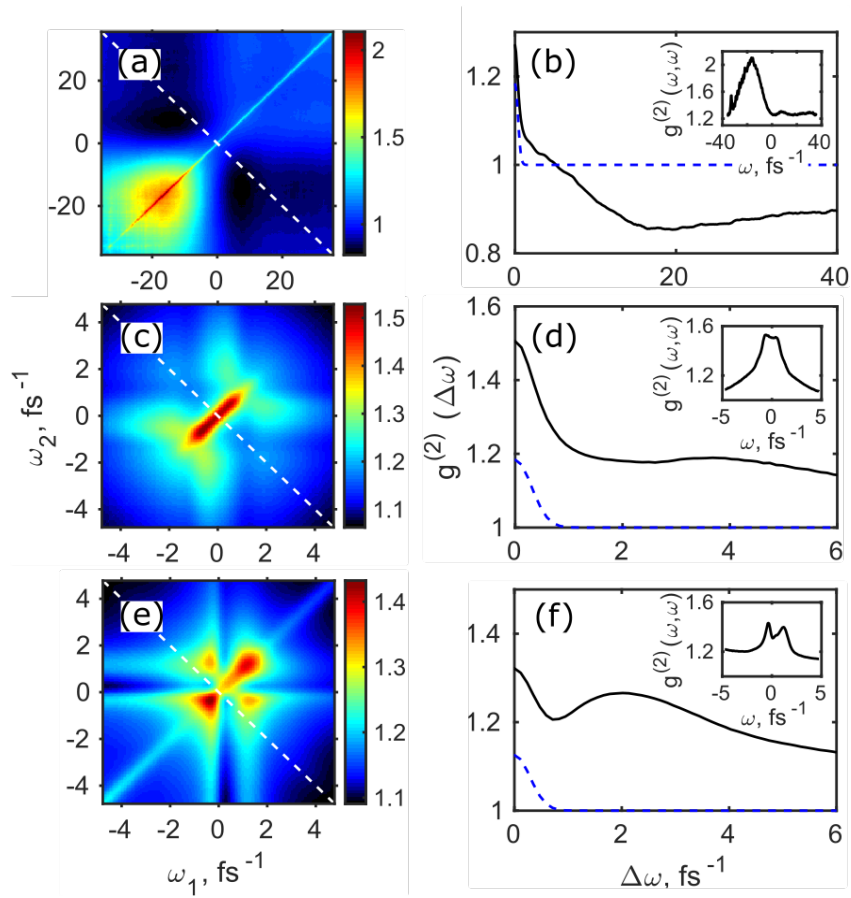

**Figure 4** (a, c, e) Spectral second order correlation function analysis at the 180 pC bunch charge. (b, d, f) Correlation functions g^((2)) (Δω) (black lines) shown along the white dashed lines in (a, c, e). Blue dashed lines show the fit of the central peak in (a, c, e). (a,b) SASE radiation, (c,d) monochromatic radiation, (e,f) self-seeding regime of operation. (b,d,f) One-dimensional profiles along the white dashed lines in (a,c,e) shown by black lines. All results presented in this figure correspond to the 180 pC bunch charge.

The shape of the narrow peak may provide an estimate of the lower value of the average pulse duration. Unfortunately, in this particular experiment we did not measure pulse duration by any alternative methods as, for example, using the cross-correlation method (Min *et al.*, 2019; Nam *et al.*, 2021; Ding *et al.*, 2012). We analysed the second-order correlation function $g^{(2)}(\omega_1, \omega_2)$ in frequency domain using the following expression (Vartanyants & Khubbutdinov, 2021; Khubbutdinov *et al.*, 2021)

$$g^{(2)}(\omega_1, \omega_2) = 1 + \zeta_S \, g_{in}(\omega_1, \omega_2), \tag{8}$$

where $\zeta_S$ is the degree of spatial coherence and $g_{in}(\omega_1, \omega_2)$ is the correlation function in front of the on-line spectrometer. For the Gaussian Schell-model pulses, when condition $\sigma_r \tau_c \ll 1$ is satisfied, where $\sigma_r$ is the rms value of the resolution function of the monochromator we obtain for $g_{in}(\omega_1, \omega_2)$ in Eq. (8) (Vartanyants & Khubbutdinov, 2021)

$$g_{in}(\Delta\omega) = \frac{exp\left[-\dfrac{\sigma_T^2}{1 + 4\sigma_r^2\sigma_T^2}(\Delta\omega)^2\right]}{(1 + 4\sigma_r^2\sigma_T^2)^{1/2}}, \tag{9}$$

where $\sigma_T$ is the rms of the average pulse duration, and the FWHM value can be obtained by expression $T = 2\sqrt{2ln2}\sigma_T \approx 2.355 \cdot \sigma_T$. For the spectrometer resolution satisfying condition $\sigma_r \sigma_T \ll 1$ Eq. (9) reduces to

$$g_{in}(\Delta\omega) = e^{-(\sigma_T \Delta\omega)^2}. \tag{10}$$

Taking into account resolution of the spectrometer $\sigma_r$=0.11 eV (Nam *et al.*, 2021) we determined from Eq. (9) the pulse duration for the SASE and monochromatic radiation to be from 6 fs to 9 fs depending on operation conditions (see Table 1).

In the case of self-seeding operation, the pulses were close to be transform limited with additional contribution of the SASE background. In addition, we observed a specific shape of $g^{(2)}(\omega_1, \omega_2)$ −function in the form of a "leaf" (see Fig. 4(e) and SI Fig. S10 for the bunch charge of 200 pC). We would leave the detailed analysis of the self-seeding mode till the Section IV.

### 3.3. Spatial analysis

An average spatial intensity distribution, measured in the case of SASE radiation with the 180 pC bunch charge, is shown in Fig. 2(c). As one can see from this figure there are some small artifacts and distortions that are present in this intensity distribution. These effects we attribute to imperfections of the KB mirrors. For the correlation analysis in the spatial domain we selected the region of interest (ROI) that was about $1 \times 1$ mm$^2$ (150 × 150 pixels), which is shown in Fig. 2(c) by the white dashed square.

The average spatial intensity distribution in horizontal and vertical directions obtained for all pulses at the 180 pC bunch charge and SASE operation mode is shown in Fig. 2(c), which looks similar to other bunch charges. To estimate the FWHM size of the beam, we performed fitting by the Gaussian functions in the vertical and horizontal directions (see Table 2). In all cases the beam size was on the order of 0.7 mm to one millimetre (FWHM).

**Table 2**   Results of the analysis in spatial domain. The beam size (FWHM) was determined by the direct evaluation of an averaged intensity distribution, the coherence length $L_{coh}$ was obtained from Eq. (10) in which integration was performed over the region where $g^{(2)}(\Delta r) \geq 1$, the degree of coherence was determined by Eq. (11) in which integration was performed over the region where $g^{(2)}(\Delta r) \geq 1$, the contrast value was defined as $\zeta_2(D_\omega) = g^{(2)}(r, r) - 1$ at $r = 0$ in the horizontal and vertical directions, respectively. The values of different parameters measured in the linear mode of operation are also provided in the Table.

| Bunch charge | 120 pC | | | 180 pC | | | 200 pC | | |
|---|---|---|---|---|---|---|---|---|---|
| Operation mode | SASE radiation | Mono-chromatic radiation | Self-seeding radiation | SASE radiation | Mono-chromatic radiation | Self-seeding radiation | SASE radiation | Mono-chromatic radiation | Self-seeding radiation |
| Horizontal direction, x | | | | | | | | | |
| Average beam size (FWHM), mm | 0.85±0.02 /1.02±0.01 (L) | 0.77±0.01 /0.95±0.01 (L) | 0.70±0.01 | 0.74±0.00 | 0.75±0.01 | 0.72±0.01 | 0.83±0.01 | 0.79±0.02 | 0.73±0.00 /0.84±0.00 (L) |
| Coherence length (rms), mm | 0.18±0.02 /0.22±0.02 (L) | 0.46±0.04 /0.67±0.06 (L) | 0.51±0.06 | 0.41±0.04 | 0.9±0.08 | 0.71±0.06 | 0.5±0.04 | 0.61±0.05 | 0.7±0.06 / 0.8±0.07 (L) |
| Degree of coherence (ζ), % | 51.2±0.90 /31.6±0.1 (L) | 76.0±3.9 /80.2±1.6 (L) | 79.0±2.9 | 60.6±1.2 | 80.6±2.7 | 80.4±1.3 | 68.5±1.6 | 78.6±6.0 | 84.2±0.5 /91.9±0.9 (L) |
| Contrast | 0.06±0.01 /0.02±0.01 (L) | 0.29±0.03 /0.21±0.03 (L) | 0.05±0.01 | 0.02±0.00 | 0.42±0.08 | 0.19±0.01 | 0.04±0.00 | 0.35±0.07 | 0.22±0.02 /0.13±0.02 (L) |
| Vertical direction, y | | | | | | | | | |
| Average beam size (FWHM), mm | 0.52±0.00 /0.61±0.00 (L) | 0.56±0.00 /0.67±0.01 (L) | 0.45±0.00 | 0.48±0.00 | 0.52±0.01 | 0.46±0.01 | 0.52±0.00 | 0.54±0.01 | 0.45±0.00 /0.50±0.00 (L) |
| Coherence length (rms), mm | 0.49±0.04 /0.33±0.03 (L) | 0.46±0.04 /0.45±0.04 (L) | 0.25±0.02 | 0.28±0.02 | 0.68±0.06 | 0.59±0.05 | 0.43±0.04 | 0.61±0.05 | 0.6±0.05 /0.6±0.05 (L) |

| Degree of coherence ($\zeta$), % | 70.8±2.2 /84.3±3.5 (L) | 84.0±0.8 /90.4±0.9 (L) | 68.3±2.0 | 72.9±0.9 | 83.3±0.8 | 76.8±1.7 | 68.2±0.8 | 81.6±1.3 | 80.5±0.6 /88.0±1.0 (L) |
|---|---|---|---|---|---|---|---|---|---|
| Contrast | 0.04±0.01 /0.01±0.00 (L) | 0.26±0.02 /0.18±0.02 (L) | 0.06±0.01 | 0.01±0.00 | 0.40±0.07 | 0.20±0.01 | 0.04±0.00 | 0.34±0.06 | 0.23±0.02 /0.14±0.02 (L) |

The second-order correlation analysis in this work was performed in the following way. We projected intensities for each pulse in the vertical and horizontal directions as $I(x) = \int_{-\infty}^{\infty} I(x,y)dy$ and $I(y) = \int_{-\infty}^{\infty} I(x,y)dx$ (see Fig. 2(c)). Next, we correlated these projected intensities according to

$$\mathrm{g}^{(2)}(x_1, x_2) = \frac{\langle I(x_1 - x_0)I(x_2 - x_0)\rangle}{\langle I(x_1 - x_0)\rangle\langle I(x_2 - x_0)\rangle}, \tag{11}$$

where $x_0$ is a centre of mass of projected intensity distribution and similar in the vertical direction. The results of intensity correlation analysis in the horizontal and vertical directions for all operation conditions with the 180 pC bunch charge are presented in Fig. 5. (see SI Figs. S13 – S20 for the other bunch charges). The intensity correlation functions determined along the white dashed lines are also shown in Fig. 5. We observed that in SASE operation regime we have two maxima along the diagonal in the bottom left and top right corners and the minimum in the middle. This is a typical behaviour of the $g^{(2)}$-function in the case of the positional jitter (Gorobtsov *et al.*, 2017*b*). In the case of monochromatic radiation and self-seeding regime of operation we observed that in most of the cases the maximum of distribution of the $g^{(2)}$-function is shifted from the centre. This effect may be due to the presence of two spatially separated beams in the intensity distribution (Gorobtsov *et al.*, 2017*b*).

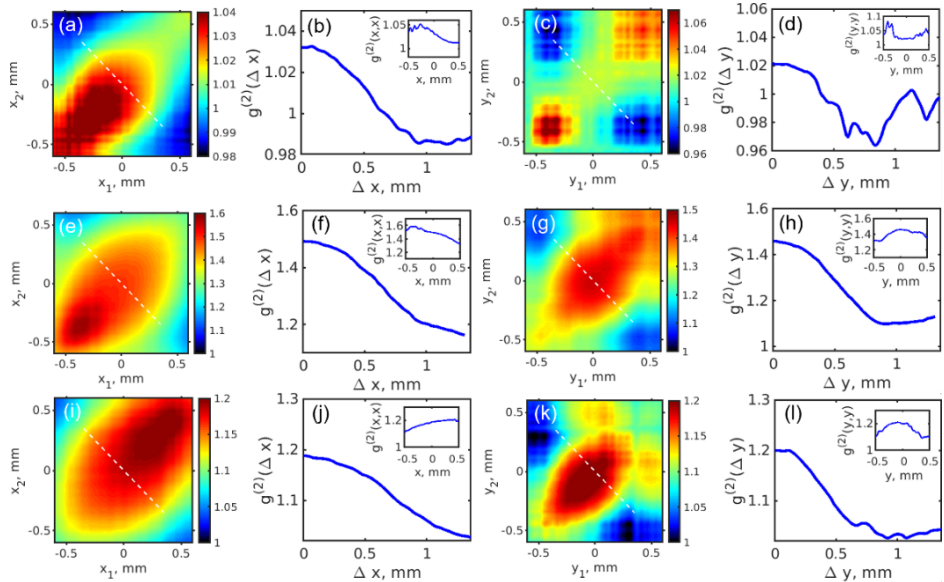

**Figure 5** Intensity correlation functions $g^{(2)}(x_1, x_2)$ (a,e,i) and $g^{(2)}(y_1, y_2)$ (c,g,k) measured in the horizontal and vertical directions, respectively. Profiles of the $g^{(2)}(\Delta x)$ (b,f,j) and $g^{(2)}(\Delta y)$ (d,h,l)

functions taken along the white dashed lines shown in panels (a,e,i) and (c,g,k), respectively. In the inset the corresponding autocorrelation functions $g^{(2)}(x, x)$ and $g^{(2)}(y, y)$ taken along the diagonal lines of $g^{(2)}$- functions are shown. (a-d) SASE radiation, (e-h) monochromatic radiation, and (i-l) self-seeding regime of operation. All results presented in this figure correspond to the 180 pC bunch charge.

To obtain the values of coherence length $L_{coh}$, we extracted one-dimensional profiles along the white dashed lines shown in Fig. 5 and determined $L_{coh}$ as variance values of these profiles (Khubbutdinov et al., 2021)

$$L_{coh}^2 = 2 \cdot \frac{\int \Delta x^2 \left|g^{(1)}(\Delta x)\right|^2 d(\Delta x)}{\int \left|g^{(1)}(\Delta x)\right|^2 d(\Delta x)} = 2 \cdot \frac{\int (\Delta x)^2 (g^{(2)}(\Delta x) - 1)\, d(\Delta x)}{\int [g^{(2)}(\Delta x) - 1] d(\Delta x)} \qquad (12)$$

and similar in the vertical direction. Eq. (12) gives an exact result for the Gaussian distribution of the first-order correlation function $g^{(1)}(\Delta x) = exp\left[-(\Delta x)^2 / 2L_{coh}^2\right]$ and integration from zero to infinity. The values of coherence length are summarized in Table 2. Here we should note, that by deriving Eq. (12) we assume that Eq. (2) is valid. At the same time Eq. (2) was obtained in the conditions of chaotic radiation. This last assumption, as we know from our previous research (Singer et al., 2013; Gorobtsov et al., 2017b; Gorobtsov, Mukharamova, Lazarev, Chollet, Zhu, Feng, Kurta, Meijer, Williams, Sikorski, & others, 2018; Khubbutdinov et al., 2021), is valid for SASE radiation without and with the monochromator, but we should be careful by applying this approach to self-seeding case. As we will discuss later, the contrast of $g^{(2)}$ −function is higher than one also for self-seeding mode of operation. However, we know that for fully coherent radiation the contrast should be equal to one (Gorobtsov, Mercurio et al., 2018). So, finally, we applied the same Eq. (12) to estimate the coherence length also for self-seeding mode of operation.

The obtained coherence length values (see Table 2) have to be compared with the rms values of the average beam size for each operation condition and evaluation direction. Such comparison shows that in the case of SASE radiation the coherence length is on the order of rms values of the average beam size and in the case of the monochromatic and self-seeding operation the coherence length exceeds the rms values of the average beam size. From this we can deduce that the degree of coherence is already high in SASE regime of operation and is substantially higher in the case of monochromatic and self-seeding operation conditions.

Next, the degree of spatial coherence $\zeta_s$ in each transverse direction was determined. The degree of spatial coherence for chaotic source can be determined according to the following equation (Gorobtsov et al., 2017b)

$$\zeta_S = \frac{\int |W(x_1, x_2)|^2 \, dx_1 dx_2}{(\int I(x) dx)^2} = \frac{\int |g^{(1)}(x_1, x_2)|^2 I(x_1) I(x_2) dx_1 dx_2}{(\int I(x) dx)^2}$$

$$= \frac{1}{\zeta_2(D_\omega)} \frac{\int [g^{(2)}(x_1, x_2) - 1] I(x_1) I(x_2) dx_1 dx_2}{(\int I(x) dx)^2}, \quad (13)$$

where $W(x_1, x_2)$ is the cross-spectral density function. The degree of spatial coherence took values in the range from 50% to 81% depending on the operation regime and electron bunch charge (see Table 2). In the SASE operating regime, the degree of spatial coherence was in most cases in the range from 50% to 70%. For the monochromatic regime of operation, the degree of spatial coherence was on the order of 80% and not depending strongly on the bunch charge. For the self-seeding regime, it was also on the order of 80% as in the monochromatic case and was slightly growing with the bunch charge. The degree of spatial coherence for Gaussian Schell-model may be determined as well from the following equation (Vartanyants & Singer, 2010, 2020)

$$\zeta_S = \left[ 1 + 4 \left( \sigma_I / L_{coh} \right)^2 \right]^{-1/2}, \quad (14)$$

where $\sigma_I$ is the rms value of the averaged intensity distribution and $L_{coh}$ is the coherence length.

The contrast values $\zeta_2(D_\omega)$ (see Eq. (2)) for all operating conditions were deduced directly from the $g^{(2)}$-function as $\zeta_2(D_\omega) = g^{(2)}(\mathbf{r}, \mathbf{r}) - 1$ at $\mathbf{r} = 0$ in the horizontal and vertical directions (see Table 2). These values of the contrast are directly related to the degree of coherence in spectral domain (Vartanyants & Khubbutdinov, 2021). In the SASE operation regime, the contrast values were from 1% to 6% depending on the bunch charge and evaluation direction. This small values concord well with an estimate for the contrast given by the following relation $\zeta_2(D_\omega) \approx 1/M_t$, where $M_t$ is the number of longitudinal modes. Our estimate of the number of longitudinal modes $M_t$ for the SASE radiation shows that it is quite high and is about 40 modes for the bunch charges of 120 pC and 200 pC (see SI Table S3). At the same time, we got quite high number of modes ~100 for the 180 pC bunch charge, that provides the low value of the contrast. As soon as number of modes is substantially decreased for the monochromatic operation (3 to 5 modes) (see SI Table S3), we expect an increase in the values of contrast for these operation conditions of the PAL-XFEL. As it follows from Table 2 the contrast values for the monochromatic radiation are in the range of 25% to 40% that match well to the estimated number of modes. Interestingly, for the self-seeding mode of operation the contrast values are in the range from 5% to 22% depending on the bunch charge. They are obviously lower than in the monochromatic case but are still not zero. We are planning to discuss all these observations in the next section.

## 4. Discussion

### 4.1. Degree of coherence

As a result of our HBT analysis, we obtained a high degree of coherence in the range of 50% to 70% for the SASE radiation in the spatial domain. The degree of coherence in the monochromatic and self-seeding operation regimes was even higher and was in the range of 75% to 85%. It is interesting to note, that the degree of coherence in the monochromatic and self-seeding operation regimes were quite similar.

## 4.2. Self-seeding operation mode

Now we turn to our basic question that was formulated in the beginning of this work: whether radiation in the self-seeding mode is fully coherent or rather has chaotic nature. To address this question, we turn our attention to the results obtained for the contrast of the spatial analysis. From our analysis we observed quite low contrast in the case of SASE radiation. This is expected, due to large amount of temporal modes $M_t$ present in each XFEL pulse. As number of modes is reduced by applying a monochromator, we obtain a significant increase in the contrast values. However, when we turn to the self-seeding mode of operation, results are quite different from the previous one. At the 120 pC bunch charge, we observe low values of contrast about 5%, indicating that at this bunch charge radiation is rather coherent (compare with the results obtained at externally seeded FEL FERMI (Gorobtsov, Mercurio *et al.*, 2018)). At the same time at the 180 pC and 200 pC bunch charges we observed that the contrast values are about 20%, that is lower than in the monochromatic case, but sufficiently larger than in the SASE case. From these results we can conclude, that in the case of self-seeding, radiation is in a mixed state: it is not fully coherent, but it is also not fully chaotic. The balance between these competing terms may be different depending on a specific tuning of the PAL-XFEL machine for this particular experiment.

## 4.3. Pulse duration

In addition, from our HBT analysis in the frequency domain we obtained comparably short pulse durations on the range of 6 fs to 9 fs, which were substantially shorter than reported earlier (Kang *et al.*, 2019; Yun *et al.*, 2019). There may be several reasons for that. For example, our results do not take into account broadening of the spectrum due to frequency chirp effects or the electron bunch compression factor (Krinsky & Li, 2006). If the electron beam is chirped this will bring in turn to broadening of the spectrum of the generated radiation. As it was shown in our previous work (Khubbutdinov *et al.*, 2021), the frequency chirp effects could bring to a substantial lower value of the pulse durations from the HBT analysis.

## 4.4. Simulations

In order to better understand some statistical features of the radiation produced by PAL-XFEL facility and revealed by our HBT analysis, we performed some additional simulations, where we used an approach based on Ref. (Pfeifer *et al.*, 2010) (see also (Khubbutdinov *et al.*, 2021)). The stochastic

XFEL radiation in the time-frequency domain with $5 \cdot 10^3$ pulses was generated by this method for each particular simulation case. For the initial simulation, the average spectrum was considered to be Gaussian and centred at the frequency $\omega_0$, corresponding to the resonant energy of $E_0 = 10$ keV. The spectral width was considered to be $\Delta E_{FWHM} = 10$ eV as in SASE radiation in our experiment. The profile of the pulse in time domain was considered to be Gaussian with the pulse duration $T_{FWHM} = 5$ fs. Results of these simulations are shown in Fig. 6. Typical single shot simulated spectra and an averaged spectrum, as well as an autocorrelation function averaged over the individual spectral lines, are shown in Fig. 6(a,b). The ACF analysis showed the FWHM size of the average spectrum 10 eV (as initially considered in the simulation), and the FWHM of single spectral spike was 0.4 eV (similar to SASE radiation case in our experiment). Analysing variation of the integrated spectral intensity distribution, we determined the number of modes, present in the simulated SASE spectrum to be about $M=28$. The second-order intensity correlation function of the simulated spectra $g^{(2)}(\omega_1,\omega_2)$ is shown in Fig. 6(c). The cut of this distribution along the diagonal line, shown by the white dashed line in Fig. 6(c) is presented in Fig. 6(d). This distribution, $g^{(2)}(\Delta\omega)$, was fitted according to Eqs. (8) and (10) and provided the initial pulse duration of 5 fs.

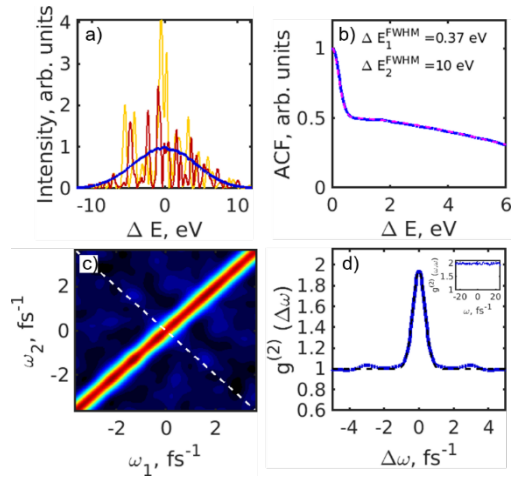

**Figure 6** Initial simulation in the spectral domain. (a) Typical single shot simulated spectra and an averaged spectrum (blue line). (b) Autocorrelation function of the individual spectral lines averaged over $5 \cdot 10^3$ pulses (blue solid line) and the fit with the two Gaussian functions (magenta dashed line). (c) Intensity correlation function $g^{(2)}(\omega_1,\omega_2)$ of the simulated spectra. (d) Intensity correlation function $g^{(2)}(\Delta\omega)$ (blue line) taken along the white dashed line in (c) and its fit (black dashed line) with Eq. (9). In the inset the profile of the $g^{(2)}(\omega,\omega)$-function along the diagonal in (c) is shown.

Since the FEL is a complicated machine, many instabilities may arise during the electron bunch acceleration and radiation amplification process. Results of such instabilities can manifest themselves, for example, in the resonant energy jitter. To study this energy jitter on the $g^{(2)}$-correlation functions, the resonant energy of 10 keV was allowed to have variations of 5 eV (FWHM) photon energy according to Gaussian distribution (see Fig. 7). As a result of these simulation we observed that the $g^{(2)}(\Delta\omega)$-correlation function along the anti-diagonal line went below one and at the same time the $g^{(2)}(\omega,\omega)$-correlation function along the diagonal line showed an increase in intensity (see Fig. 7(e,f)).These both effects were similar to those observed in our experiment (compare to Fig. 4(a,b)). The above-mentioned features indicate the presence of the energy jitter effects in our experiment.

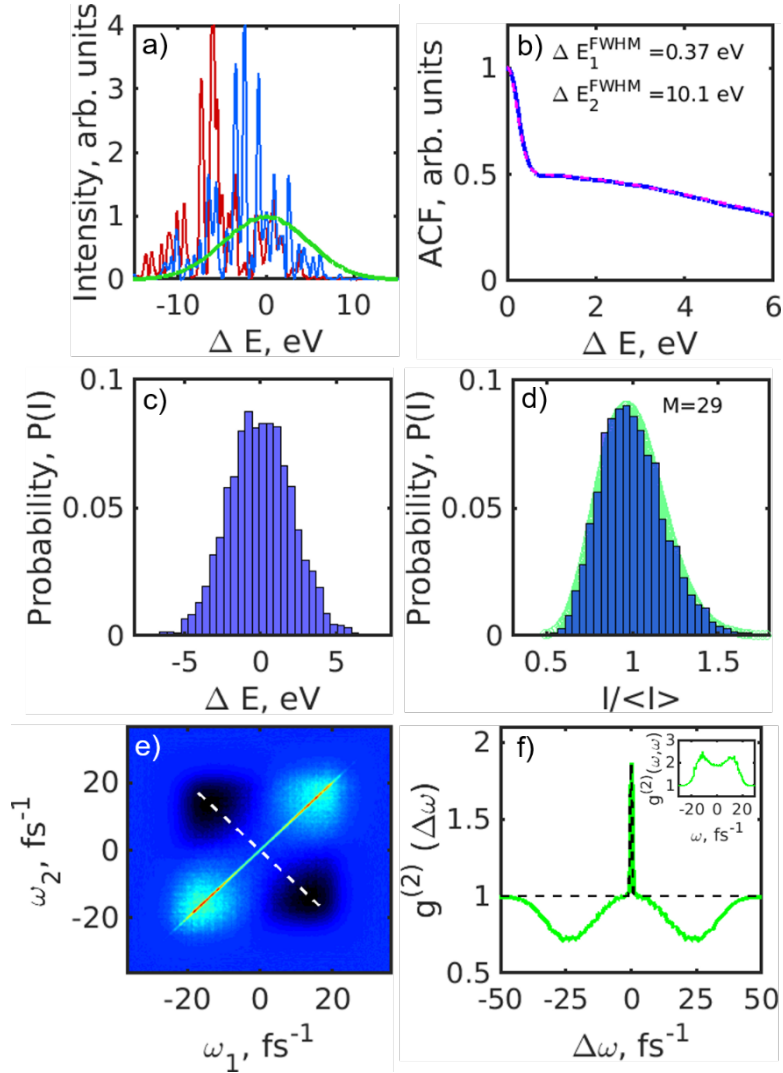

**Figure 7** Spectral analysis simulation with the energy jitter of 5 eV (FWHM). (a) Typical single shot simulated spectra and an averaged spectrum (green line). (b) Autocorrelation function of the individual spectral lines averaged over $5 \cdot 10^3$ pulses (blue solid line) and the fit with the two Gaussian functions (magenta dashed line). (c) Histogram of the resonant energy distribution. (d) Histogram of the spectral pulse intensity distribution (blue). The green background corresponds to the gamma probability distribution function with the number of modes *M*=29. (e) Second-order intensity correlation function $g^{(2)}(\omega_1,\omega_2)$ of the simulated spectra. (f) Second-order intensity correlation function $g^{(2)}(\Delta\omega)$ taken along the white dashed line in (c) and its fit (black dashed line) with Eq. (9). In the inset the profile of $g^{(2)}(\omega,\omega)$ function along the diagonal in (e) is shown.

Along with the energy jitter, it may be also the pulse duration jitter from pulse to pulse that may affect the observed $g^{(2)}(\omega_1, \omega_2)$-correlation function. To study this effect, we simulated pulses with one femtosecond (rms) variations from pulse to pulse following Gaussian distribution. As a result of these simulations, a small 'bump' in the distribution of the g(2)($\Delta\omega$)-correlation function was observed (see Fig. 8(f)), that was similar to our experimental results (see Fig. 4(b)). The presence of such a broadening in the correlation functions obtained from our experimental data may indicate a possible pulse duration jitter at PAL-XFEL facility.

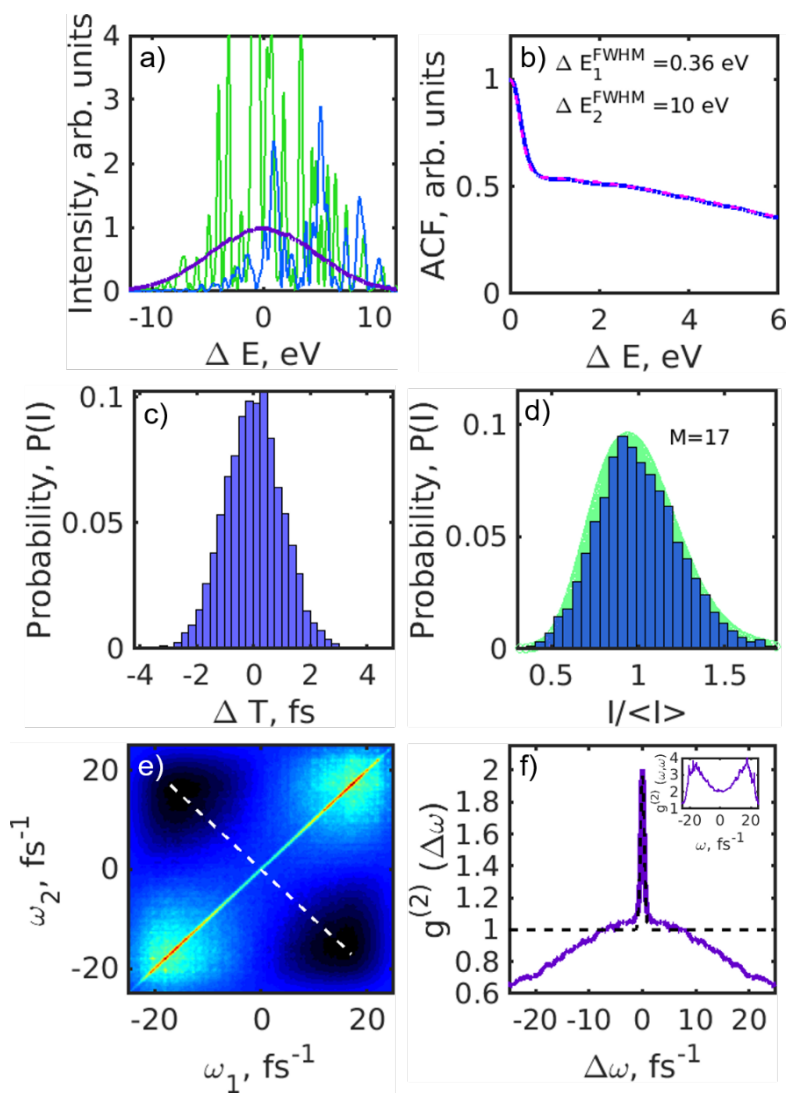

**Figure 8** Spectral analysis simulation with the energy jitter of 5 eV (FWHM) and additional pulse duration jitter of 1 fs (rms). (a) Typical single shot simulated spectra and an averaged spectrum (blue line). (b) Autocorrelation function of individual spectral lines averaged over $5 \cdot 10^3$ pulses (blue line) and the fit with the two Gaussian functions (magenta dashed line). (c) Histogram of the pulse duration distribution. (d) Histogram of the spectral pulse intensity distribution (blue). The green background corresponds to the gamma probability distribution function with the number of modes $M$=17. (e) Second-order intensity correlation function $g^{(2)}(\omega_1,\omega_2)$ of the simulated spectra. (f) Second-order

intensity correlation function $g^{(2)}(\Delta\omega)$ taken along the white dashed line in (c) and its fit (black dashed line) with Eq. (9). Additional features around $\Delta\omega=0$ are well seen in this panel. In the inset the profile of $g^{(2)}(\omega,\omega)$ function along the diagonal in (e) is shown.

We also simulated results of monochromatic radiation on $g^{(2)}$-function. For that, for generated pulses we applied a bandwidth of $\Delta E=1.9$ eV in frequency domain (see Fig. 9). In the distribution of modes we obtained only two modes that considerably contribute to the result. For the $g^{(2)}(\omega_1, \omega_2)$-correlation function we obtained result shown in Fig. 9(c) that is similar to our experimental result for monochromatic case (see Fig. 4(c)).

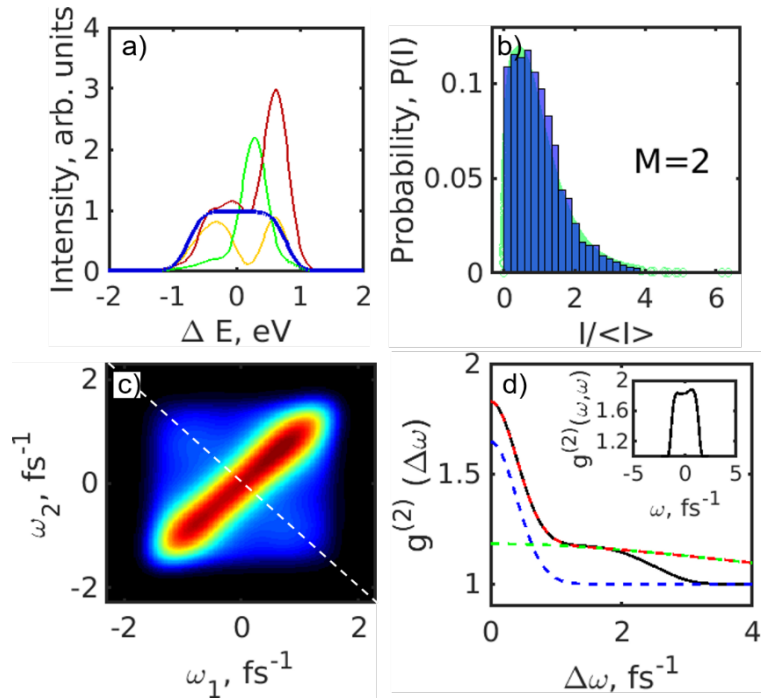

**Figure 9** Spectral analysis simulation with the monochromator installed in the beamline. (a) Simulated profiles of single pulse intensities and an averaged spectrum (blue line) with the bandwidth of 1.9 eV. (b) Histogram of intensities of single pulses obeying Gamma function distribution (green background) with the number of modes $M=2$. (c) Second-order intensity correlation function $g^{(2)}(\omega_1,\omega_2)$ of the simulated spectra with the monochromator. (d) Second-order intensity correlation function $g^{(2)}(\Delta\omega)$ (black curve) taken along the white dashed line in (c) and its fit (red dashed line) with Eq. (9). In the inset the profile of $g^{(2)}(\omega,\omega)$ function along the diagonal in (c) is shown.

Next, we turned to simulation of self-seeded pulses. We used the same approach and fixed the pulse duration to be about $T=5$ fs and, at the same time, reduced the bandwidth of the generated pulses in frequency domain ($\Delta E=0.4$ eV (FWHM)) until we get a single mode distribution (see Fig. 10(g)). It is interesting to note that each pulse in this simulation was having a varying phase both in the energy and time domains that was random from pulse to pulse (see Fig. 10(c)). Then the pulses were modified in time domain by putting a constant value to the phases of each pulse and allow these phases to change randomly from pulse to pulse (see Fig. 10(d)). To our surprise in this case we obtained the shape of the $g^{(2)}(\omega_1, \omega_2)$-correlation function in the form of the leaf (see Fig. 10(e)), similar to our results for self-seeding operation mode (see Fig 4(e)). We also noticed that the distance between two maxima along the anti-diagonal line depends on the pulse duration. To analyse this in detail we plotted this distance in frequency $\Delta\omega$ as a function of the pulse duration and obtained the curve shown in Fig. 10(h). From that curve we identified that for the distance between two maxima determined from our experiment ($\Delta\omega$=3.5 fs$^{-1}$) we obtain a pulse duration for the self-seeding operation mode about 7 fs.

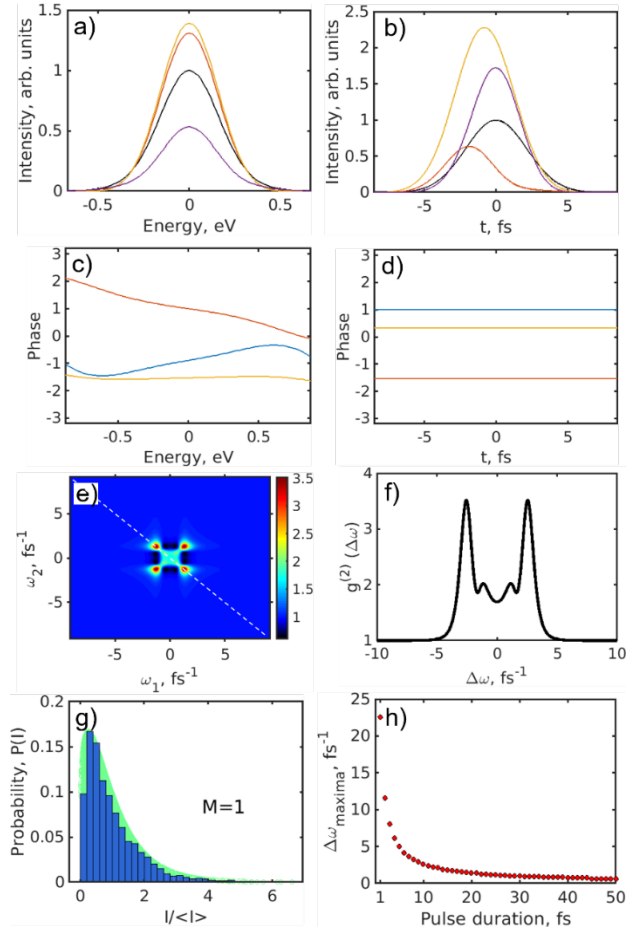

**Figure 10**    Spectral simulations for the seeded beam. (a-d) An example of generated single pulses in (a) energy domain with the bandwidth 0.4 eV and (b) time domain with the pulse duration $T$=5 fs (FWHM).  (c,d) The corresponding phases for these pulses in energy domain (c) and in time domain (d). The phases in time domain were put to a constant value corresponding to phase at $T$=0. (e) Second-order intensity correlation function $g^{(2)}(\omega_1,\omega_2)$ of the simulated spectra of the seeded beam. (f) Second-order intensity correlation function $g^{(2)}(\Delta\omega)$ (black curve) taken along the white dashed line in (e). (g) Histogram of intensities of single pulses obeying Gamma function distribution (green background) with the number of modes $M$=1. (h) Distance between two maxima $\Delta\omega_{maxima}$ in $g^{(2)}(\Delta\omega)$-correlation function in (f) as a function of pulse duration.

## 5. Summary

In summary, the statistical analysis was performed for the characteristics of hard X-ray at PAL-XFEL through the HBT interferometer technique. In particular, the information on average energy distribution, coherence time, and pulse duration could be obtained by spectral analysis, and the information on beam size, coherence length, degree of coherence can be obtained by the spatial intensity analysis under various conditions (SASE, monochromatic, and self-seeded radiation at the 120 pC, 180 pC, and 200 pC bunch charge).

The results of this experiment not only allow us to understand the present performance of the PAL-XFEL but will be an important factor for a facility upgrades in the future.

**Acknowledgements**     Y.Y.K., R.Kh., J.C. and I.A.V. are thankful to E. Weckert for the support of the project. The authors acknowledge careful reading of the manuscript by G. Mercurio.

Grunewald, S., Grzegory, P., Feng, G., Guler, H., Gusev, G., Gutierrez, J. L., Hagge, L., Hamberg, M., Hanneken, R., Harms, E., Hartl, I., Hauberg, A., Hauf, S., Hauschildt, J., Hauser, J., Havlicek, J., Hedqvist, A., Heidbrook, N., Hellberg, F., Henning, D., Hensler, O., Hermann, T., Hidvégi, A., Hierholzer, M., Hintz, H., Hoffmann, F., Hoffmann, M., Hoffmann, M., Holler, Y., Hüning, M., Ignatenko, A., Ilchen, M., Iluk, A., Iversen, J., Iversen, J., Izquierdo, M., Jachmann, L., Jardon, N., Jastrow, U., Jensch, K., Jensen, J., Jeżabek, M., Jidda, M., Jin, H., Johannson, N., Jonas, R., Kaabi, W., Kaefer, D., Kammering, R., Kapitza, H., Karabekyan, S., Karstensen, S., Kasprzak, K., Katalev, V., Keese, D., Keil, B., Kholopov, M., Killenberger, M., Kitaev, B., Klimchenko, Y., Klos, R., Knebel, L., Koch, A., Koepke, M., Köhler, S., Köhler, W., Kohlstrunk, N., Konopkova, Z., Konstantinov, A., Kook, W., Koprek, W., Körfer, M., Korth, O., Kosarev, A., Kosiński, K., Kostin, D., Kot, Y., Kotarba, A., Kozak, T., Kozak, V., Kramert, R., Krasilnikov, M., Krasnov, A., Krause, B., Kravchuk, L., Krebs, O., Kretschmer, R., Kreutzkamp, J., Kröplin, O., Krzysik, K., Kube, G., Kuehn, H., Kujala, N., Kulikov, V., Kuzminych, V., La Civita, D., Lacroix, M., Lamb, T., Lancetov, A., Larsson, M., Le Pinvidic, D., Lederer, S., Lensch, T., Lenz, D., Leuschner, A., Levenhagen, F., Li, Y., Liebing, J., Lilje, L., Limberg, T., Lipka, D., List, B., Liu, J., Liu, S., Lorbeer, B., Lorkiewicz, J., Lu, H. H., Ludwig, F., Machau, K., Maciocha, W., Madec, C., Magueur, C., Maiano, C., Maksimova, I., Malcher, K., Maltezopoulos, T., Mamoshkina, E., Manschwetus, B., Marcellini, F., Marinkovic, G., Martinez, T., Martirosyan, H., Maschmann, W., Maslov, M., Matheisen, A., Mavric, U., Meißner, J., Meissner, K., Messerschmidt, M., Meyners, N., Michalski, G., Michelato, P., Mildner, N., Moe, M., Moglia, F., Mohr, C., Mohr, S., Möller, W., Mommerz, M., Monaco, L., Montiel, C., Moretti, M., Morozov, I., Morozov, P., Mross, D., Mueller, J., Müller, C., Müller, J., Müller, K., Munilla, J., Münnich, A., Muratov, V., Napoly, O., Näser, B., Nefedov, N., Neumann, R., Neumann, R., Ngada, N., Noelle, D., Obier, F., Okunev, I., Oliver, J. A., Omet, M., Oppelt, A., Ottmar, A., Oublaid, M., Pagani, C., Paparella, R., Paramonov, V., Peitzmann, C., Penning, J., Perus, A., Peters, F., Petersen, B., Petrov, A., Petrov, I., Pfeiffer, S., Pflüger, J., Philipp, S., Pienaud, Y., Pierini, P., Pivovarov, S., Planas, M., Pławski, E., Pohl, M., Polinski, J., Popov, V., Prat, S., Prenting, J., Priebe, G., Pryschelski, H., Przygoda, K., Pyata, E., Racky, B., Rathjen, A., Ratuschni, W., Regnaud-Campderros, S., Rehlich, K., Reschke, D., Robson, C., Roever, J., Roggli, M., Rothenburg, J., Rusiński, E., Rybaniec, R., Sahling, H., Salmani, M., Samoylova, L., Sanzone, D., Saretzki, F., Sawlanski, O., Schaffran, J., Schlarb, H., Schlösser, M., Schlott, V., Schmidt, C., Schmidt-Foehre, F., Schmitz, M., Schmökel, M., Schnautz, T., Schneidmiller, E., Scholz, M., Schöneburg, B., Schultze, J., Schulz, C., Schwarz, A., Sekutowicz, J., Sellmann, D., Semenov, E., Serkez, S., Sertore, D., Shehzad, N., Shemarykin, P., Shi, L., Sienkiewicz, M., Sikora, D., Sikorski, M., Silenzi, A., Simon, C., Singer, W., Singer, X., Sinn, H., Sinram, K., Skvorodnev, N., Smirnow, P., Sommer, T., Sorokin, A., Stadler, M., Steckel, M., Steffen, B., Steinhau-Kühl, N., Stephan, F., Stodulski, M., Stolper, M., Sulimov, A.,

# Supporting information

## S1.    Data processing

In this experiment, we simultaneously measured intensity information in the spectral and spatial domains. The number of pulses analysed in this work for each bunch charge and operation condition of the PAL XFEL are listed in Table S1. In order to extract accurate information of the data, dark images of about 1000 shots were obtained in each condition, the average value was calculated, and the dark image was subtracted from each data in spectral and spatial domains. After processing of these data, a projection along the vertical direction was performed in 2D spectral detector. One example is shown in Fig. S1 (a,b). In each condition, the average profile from all pulses of the one-dimensional spectra was obtained, the zero value of $\Delta E$ was set through the fitting of Gaussian distribution, and the value of full width of half maximum (FWHM) for energy distribution was obtained (see Fig. S1 (c)).

For each pulse, a spatial image was also projected vertically and horizontally to obtain the one-dimensional profile. After that, the average value for all pulses was calculated, the zero position was set through the fitting of the Gaussian distribution in each direction, and the FWHM for the beam size was calculated (see Fig. S1 (e,f)).

**Table S1**    Number of pulses that was analysed in this work.

| Bunch charge | 120 pC | | | 180 pC | | | 200 pC | | |
|---|---|---|---|---|---|---|---|---|---|
| Operation mode | SASE radiation | Mono-chromatic radiation | Self-seeding radiation | SASE radiation | Mono-chromatic radiation | Self-seeding radiation | SASE radiation | Mono-chromatic radiation | Self-seeding radiation |
| Number of pulses | 9,833 /16,876 (L) | 11,403 /19,961 (L) | 19,938 | 11,866 | 9,513 | 19,869 | 7,940 | 8,405 | 19,781 /19,936 (L) |

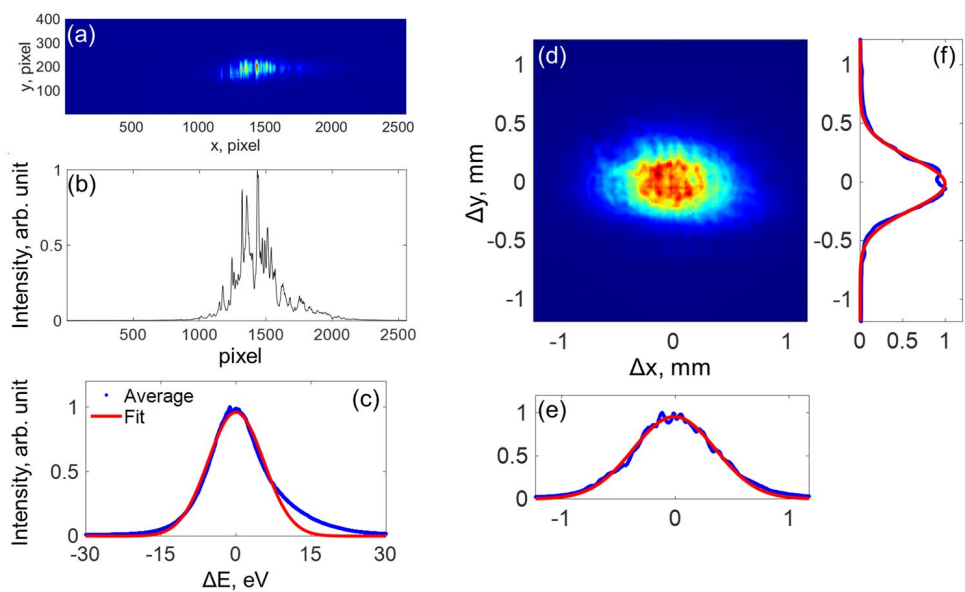

**Figure S1** Results of the spectrum and intensity distributions in spatial domain for SASE radiation at 120 pC bunch charge. From all shown images an average dark background from 1,000 pulses was subtracted. (a) Image of a single spectrum, (b) projection of this spectrum along the vertical direction, (c) average spectral intensity (blue line) and the fit with the Gaussian function (red line). (d) Average intensity at Hamamatzu detector. (e,f) Projection of this intensity in the vertical (e) and horizontal (f) directions. Blue lines are the average intensities for each projection and red lines are the fit with the Gaussian function.

## S2. Monochromator drift corrections

During the measurements for the case of SASE with the monochromator, we observed the vertical position drifts of the monochromator (see Fig. S2). So, we set the sections where the drift was happening rapidly, calculated the linear regression for each section, and then corrected it (see Fig. S2 (b,c)). This correction was performed for all monochromatic radiation modes.

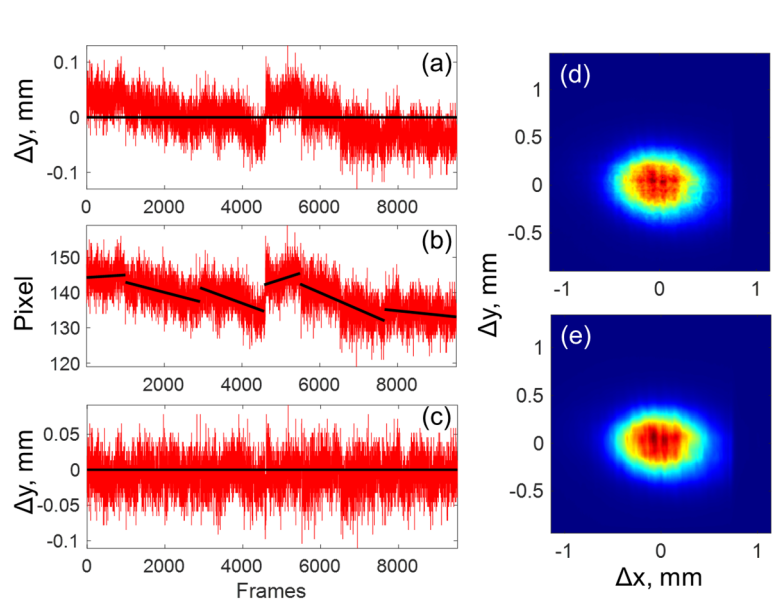

**Figure S2** Spatial corrections for the monochromator mode of operation at the 180 pC bunch case. (a) Centre of mass position along the vertical direction at each frame. (b) linear regression at each selected region. (c) Centre of mass position after correction. Average spatial intensity before correction (d) and after correction (e).

### S3.    Spectral profile analysis

The 200 pulses of the one-dimensional single spectral profiles, an average profile, and the autocorrelation function (ACF) for all operating conditions at 120 pC and 200 pC are shown in Figs. S3-S4. The analysis of the ACF determined the bandwidth of the spectrum related to the average FEL bandwidth as well as a single spike width (Khubbutdinov *et al.*, 2021). The averaged ACF was obtained by calculating the autocorrelation functions for each spectrum individually and then averaging it over all pulses, for certain conditions of operation of the PAL XFEL. To extract the FWHM values from the ACF's for all operation conditions, we performed the Gaussian function fitting. The average FWHM width of the peaks in spectra $\Delta E$ is related to the width of the peaks determined from the autocorrelation function $\Delta E_{ACF}$ as $\Delta E = \Delta E_{ACF}/\sqrt{2}$.

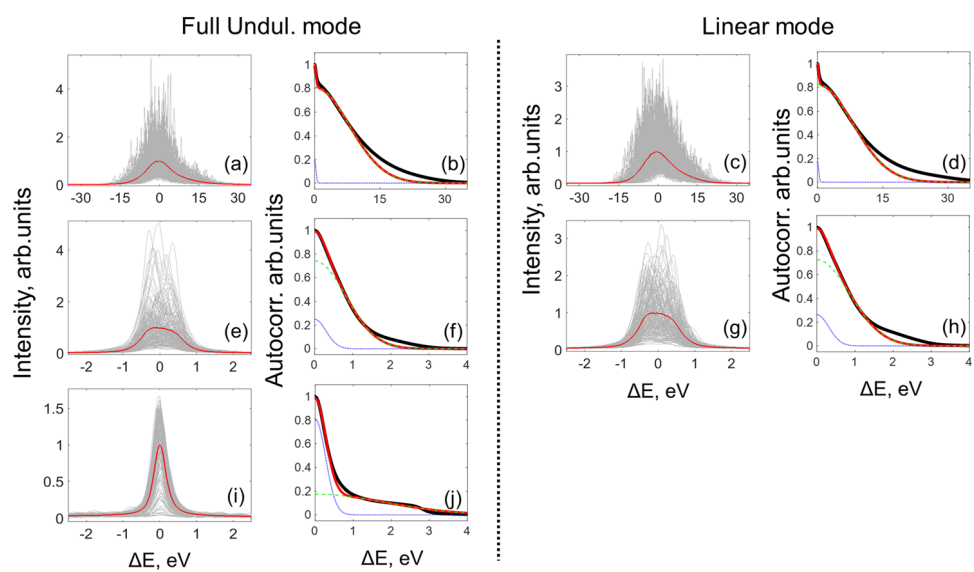

**Figure S3** (a, e, i) Spectral distribution of 200 pulses (grey lines) and an average spectrum for all pulses (red lines). (b, f, g) Average autocorrelation functions of all spectral lines (black lines), and its fit (red lines) by two Gaussian functions, broad (green dashed lines) and narrow (blue dotted lines). (a,b) SASE radiation, (e,f) SASE monochromatic radiation, (i,j) self-seeding regime of operation. (c,d) Same in SASE linear mode of operation, (g,h) Same in SASE linear mode of operation with the monochromator. All results presented in this figure correspond to the 120 pC bunch charge.

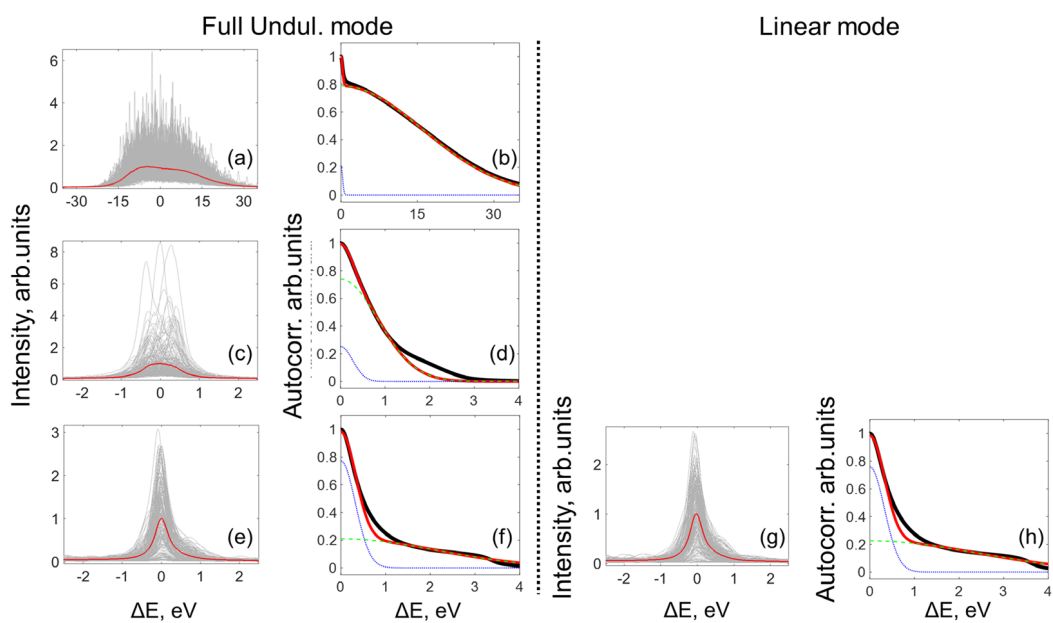

**Figure S4** (a, c, e) Spectral distribution of 200 pulses (grey lines) and an average spectrum for all pulses (red lines). (b, d, f) Average autocorrelation functions of all spectral lines (black lines), and its fit (red lines) by two Gaussian functions, broad (green dashed lines) and narrow (blue dotted lines). (a,b) SASE radiation, (e,f) SASE monochromatic radiation, (i,j) self-seeding regime of operation. (g,h)

Same in self-seeding linear mode of operation. All results presented in this figure correspond to the 200 pC bunch charge.

## S4.    An average spectral profile fit

As it was described in the main text to determine coherence time one needs to obtain an averaged spectral profile. We noticed that in the case of PAL XFEL an averaged spectral profile is well fitted by two Gaussian functions as

$$S(\omega) = A_1 exp\left[-\frac{(\omega - \omega_1^0)^2}{2\sigma_1^2}\right] + A_2 exp\left[-\frac{(\omega - \omega_2^0)^2}{2\sigma_2^2}\right], \quad \text{(S1)}$$

where $A_1$ and $A_2$ are scaling coefficients, $\omega_1^0$ and $\omega_2^0$ are the centres of each Gaussian line and $\sigma_1$ and $\sigma_2$ are their rms values. Results of such fitting are summarized in Table S2 and shown in Fig. S5. Substituting this spectral profile in Eqs. (3,4) of the main text one can obtain for the coherence time an expression given in Eq. (5) of the main text (Khubbutdinov *et al.*, 2021).

**Table S2**    Parameters of the two Gaussian functions obtained from the fit of the averaged spectrum in different operation conditions (see Eq. (S1)).

| Bunch charge | 120 pC | | | 180 pC | | | 200 pC | | |
|---|---|---|---|---|---|---|---|---|---|
| Operation mode | SASE radiation | Mono-chromatic radiation | Self-seeding radiation | SASE radiation | Mono-chromatic radiation | Self-seeding radiation | SASE radiation | Mono-chromatic radiation | Self-seeding radiation |
| $A_1$ | 0.657 /0.590(L) | 0.761 /0.741(L) | 0.744 | 0.701 | 0.533 | 0.692 | 0.469 | 0.749 | 0.646 /0.639(L) |
| $\omega_1^0$, eV | -0.941 /-1.745(L) | -0.413 /-0.435(L) | 0.004 | -0.846 | -0.548 | -0.016 | -11.497 | -0.240 | 0.005 /-0.045(L) |
| $\sigma_1$, eV | 6.081 /6.428(L) | 0.465 /0.472(L) | 0.216 | 5.925 | 0.428 | 0.215 | 7.219 | 0.554 | 0.212 /0.219(L) |
| $A_2$ | 0.354 /0.440(L) | 0.735 /0.723(L) | 0.234 | 0.316 | 0.879 | 0.301 | 0.851 | 0.442 | 0.327 /0.336(L) |
| $\omega_2^0$, eV | 4.147 /4.476(L) | 0.506 /0.474(L) | 0.050 | 5.000 | 0.307 | 0.159 | 6.538 | 0.545 | 0.110/ 0.074(L) |
| $\sigma_2$, eV | 14.149 /12.920 (L) | 0.525 /0.520(L) | 0.652 | 14.993 | 0.550 | 0.633 | 16.170 | 0.526 | 0.711 /0.754(L) |

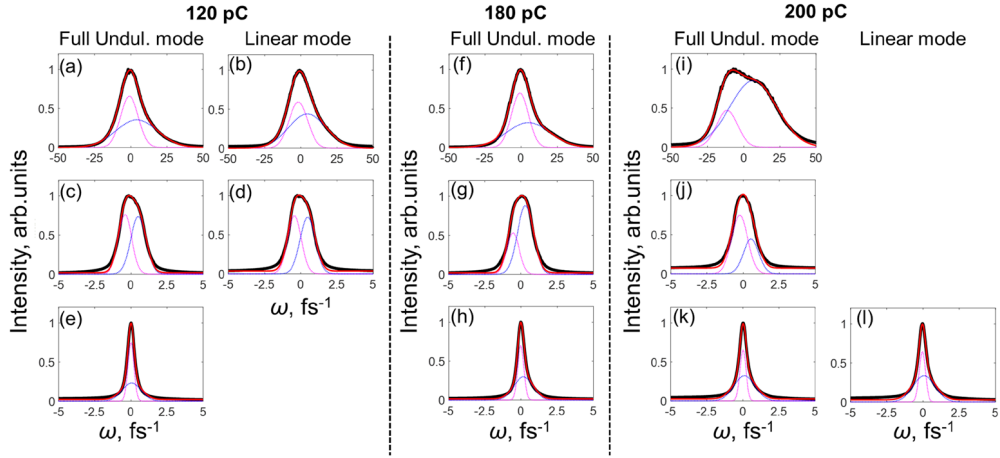

**Figure S5** Results of fitting (red line) of an averaged spectrum (black line) by two Gaussian functions (magenta and blue lines) for different operation conditions. (a, f, i) SASE mode, (b) SASE linear mode, (c, g, j) SASE monochromatic conditions, (d) SASE monochromatic linear mode, (e, h, k) self-seeding mode, and (l) self-seeding linear operation mode. Measurements were performed at the 120 pC (a – e), 180 pC (f – h), and 200 pC (i – l) bunch charges, respectively.

## S5. Intensity histograms and calculation of the number of modes

To analyse the statistical behaviour of FEL radiation we looked on the histograms of the pulse intensity distributions. These histograms were evaluated for all operation conditions from spectral and spatial measurements of PAL XFEL source and are presented in Figs. S6- S8. The histograms were compared with the Gamma probability distribution function which describes statistical behaviour of the FEL SASE radiation in the linear regime of operation (Saldin *et al.*, 2000)

$$p\left(\frac{I}{\langle I \rangle}\right) = \frac{M^M}{\Gamma(M)}\left(\frac{I}{\langle I \rangle}\right)^{M-1}\frac{I}{\langle I \rangle}exp\left(-M\frac{I}{\langle I \rangle}\right), \tag{S2}$$

where $I$ is the integrated intensity for a single pulse, $\langle I \rangle$ is the average intensity from all pulses and $M$ is the number of modes.

According to the FEL theory (Saldin *et al.*, 2000), the number of modes M is inversely proportional to the normalized dispersion of the intensity distribution

$$M = \frac{\langle I \rangle^2}{\sigma_I^2}, \tag{S3}$$

where $\sigma_I$ is the standard deviation of the intensity distribution. The results of this analysis were summarized in Table S3.

**Table S3**    Results of the mode analysis in spatial and spectral domains.

| Bunch charge | 120 pC | | | 180 pC | | | 200 pC | | |
|---|---|---|---|---|---|---|---|---|---|
| Operation mode | SASE | mono | seed | SASE | mono | seed | SASE | mono | seed |
| M Spatial domain | 37.0±4.9 /201.5±62.7 (L) | 4.8±0.5 /6.2±0.6 (L) | 26.3±3.9 | 102.6±12.7 | 3.1±0.1 | 7.0±1.0 | 40.2±3.5 | 4.2±0.4 | 5.9±0.6 /8.6±0.7 (L) |
| M Spectral domain | 28.3±3.0 /305.1±76.3 (L) | 11.7±3.0 /61.2±16.9 (L) | 27.1±1.4 | 43.4±2.8 | 11.0±2.6 | 6.3±0.4 | 64.3±6.2 | 401.1±296.9 | 5.7±0.5 /11.8±1.0 (L) |

As we can see from Figs. S6- S8 Gamma distribution is observed mostly in monochromatic regime of operation when there are few modes are contributing to the total intensity. In the SASE mode number of modes is large (about 100) and analysis by Gamma distribution is not working. In self-seeding mode pulse distribution is complicated and will need a special analysis.

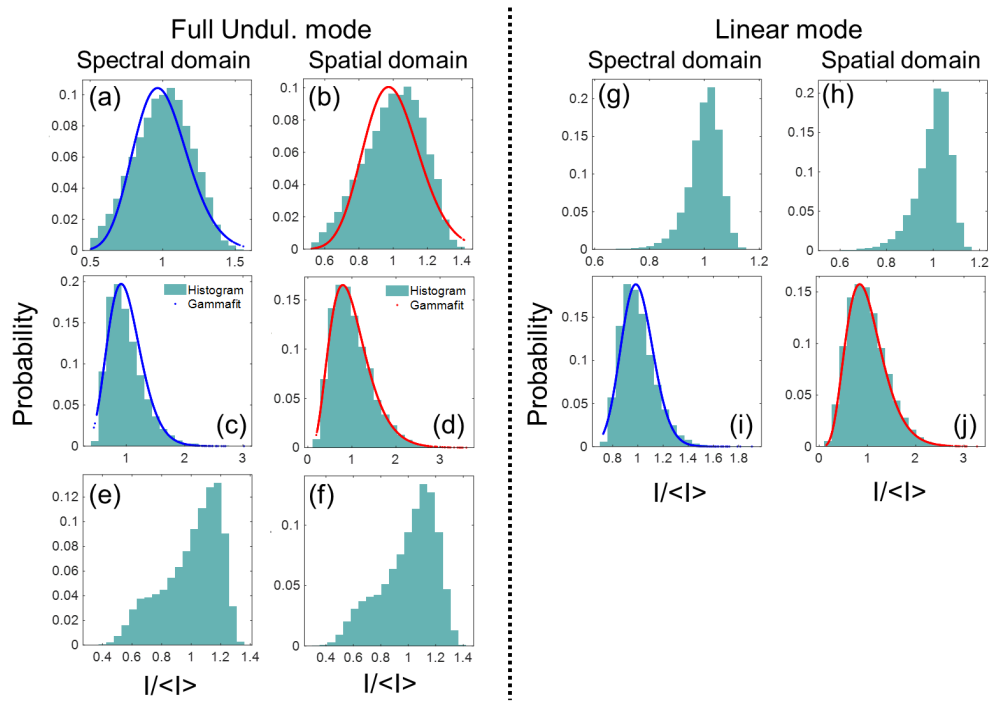

**Figure S6**  Intensity histograms in the spatial and spectral domains for operation of the PAL XFEL source at the 120 pC bunch charge. Left side: (a,b) SASE mode, (c,d) SASE monochromatic mode, and (e,f) self-seeding mode of operation. Right side, linear mode of operation: (g,h) SASE and (i,j) SASE monochromatic mode. The red and blue lines are the results of Gamma distribution fitting.

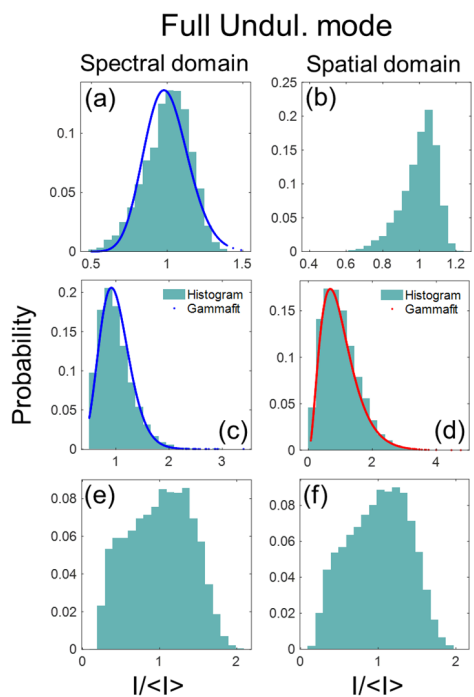

**Figure S7** Intensity histograms in the spatial and spectral domains for operation of the PAL XFEL source at the 180 pC bunch charge. (a,b) SASE mode, (c,d) SASE monochromatic mode, and (e,f) self-seeding mode of operation. The red and blue lines are the results of Gamma distribution fitting.

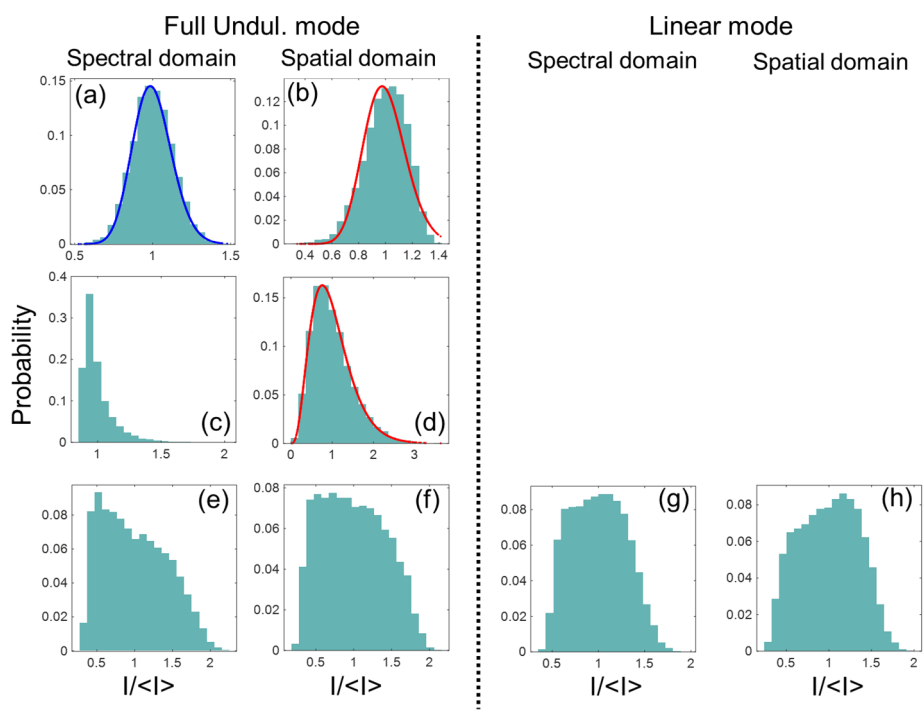

**Figure S8** Intensity histograms in the spatial and spectral domains for operation of the PAL XFEL source at the 200 pC bunch charge. Left side: (a,b) SASE mode, (c,d) SASE monochromatic mode, and (e,f) self-seeding mode of operation. Right side, linear mode of operation: (g,h) self-seeding mode of operation in linear regime. The red and blue lines are the results of Gamma distribution fitting.

## S6.    HBT analysis

### S6.1. Spectral analysis

The normalized spectral $g^{(2)}(\omega_1,\omega_2)$-correlation function has the following form

$$g^{(2)}(\omega_1,\omega_2) = \frac{<I(\omega_1-\omega_0)I(\omega_2-\omega_0)>}{<I(\omega_1-\omega_0)><I(\omega_2-\omega_0)>}, \qquad (S4)$$

where $I(\omega_1)$ and $I(\omega_2)$ are the intensities of the wave field in spectral representation, $\omega_0$ is the central frequency, and averaging denoted by brackets $<...>$ is performed over a large ensemble of different realizations of the wave field. These spectral correlation functions are presented in Figs. S9-S12 and are discussed in the main part of the paper.

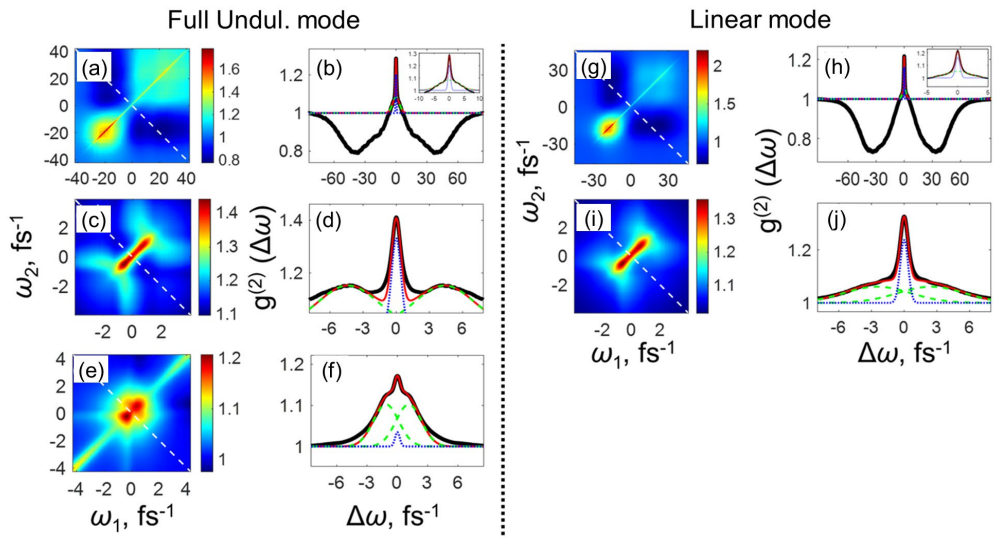

**Figure S9** Spectral $g^{(2)}$-function analysis and one-dimensional anti-diagonal profile (along white dash line) for operation of the PAL XFEL source at the 120 pC bunch charge. Left side: (a,b) SASE mode, (c,d) SASE monochromatic mode, and (e,f) self-seeding mode of operation. Right side, linear mode of operation: (g,h) SASE mode, (i,j) SASE monochromatic mode. The inset is enlarged profile at the (b,h). The pulse duration was calculated by blue dash line peak

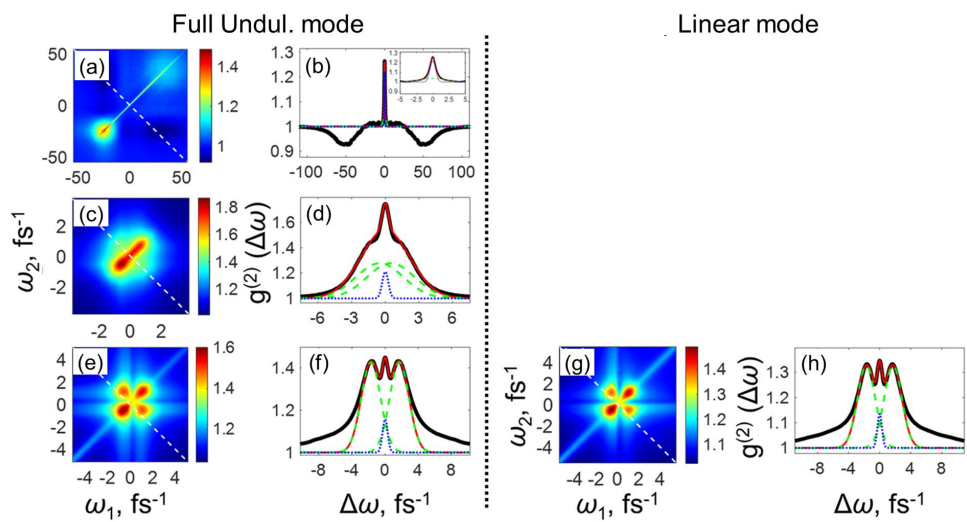

**Figure S10** Spectral $g^{(2)}$-function analysis and one-dimensional anti-diagonal profile (along white dash line) for operation of the PAL XFEL source at the 200 pC bunch charge. Left side: (a,b) SASE mode, (c,d) SASE monochromatic mode, and (e,f) self-seeding mode of operation. Right side, linear mode of operation: (g,h) self-seeding mode of operation. The inset is enlarged profile at the (b,h). The pulse duration was calculated by blue dash line peak.

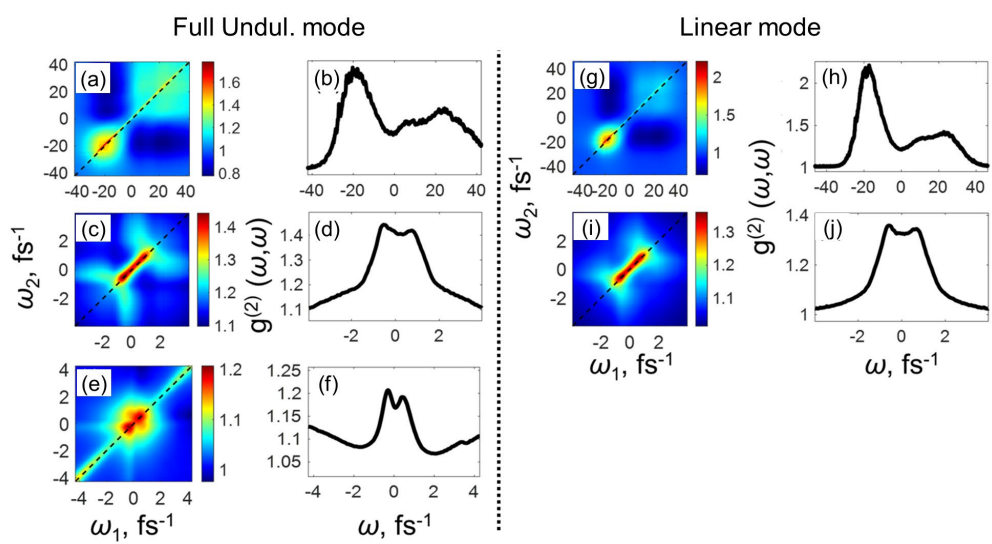

**Figure S11** Spectral $g^{(2)}$-function analysis and one-dimensional diagonal profile (along black dash line) for operation of the PAL XFEL source at the 120 pC bunch charge. Left side: (a,b) SASE mode, (c,d) SASE monochromatic mode, and (e,f) self-seeding mode of operation. Right side, linear mode of operation: (g,h) SASE mode, (i,j) SASE monochromatic mode.

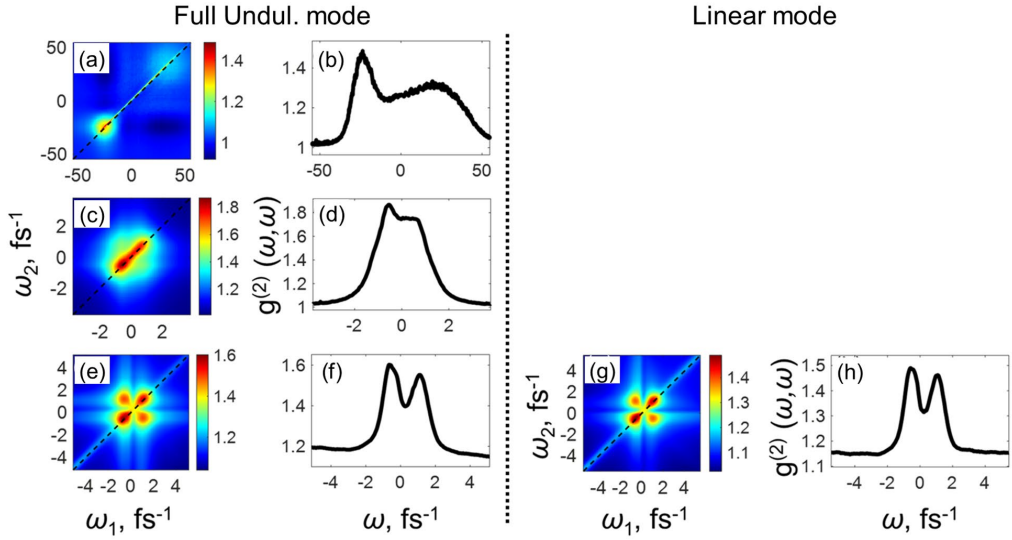

**Figure S12** Spectral $g^{(2)}$-function analysis and one-dimensional diagonal profile (along black dash line) for operation of the PAL XFEL source at the 200 pC bunch charge. Left side: (a,b) SASE mode, (c,d) SASE monochromatic mode, and (e,f) self-seeding mode of operation. Right side, linear mode of operation: (g,h) self-seeding mode of operation.

## S6.2. Spatial analysis

The normalized spatial second-order correlation function is expressed as

$$g^{(2)}(x_1, x_2) = \frac{\langle I(x_1 - x_0) I(x_2 - x_0) \rangle}{\langle I(x_1 - x_0) \rangle \langle I(x_2 - x_0) \rangle}. \tag{S5}$$

Here $I(x) = \int_{-\infty}^{\infty} I(x, y) dy$ is the projected intensity for each pulse in the vertical direction, $I(x_1)$, $I(x_2)$ are the intensities of the wave field at different positions $x_1$ and $x_2$, and averaging denoted by brackets $\langle \ldots \rangle$ is performed over a large ensemble of different realizations of the wave field. Similar to Eq. (S5) expression is valid in the vertical direction. These spatial correlation functions analysed for all operation conditions at PAL XFEL are presented in Figs. S13-S20 and are discussed in the main part of the paper.

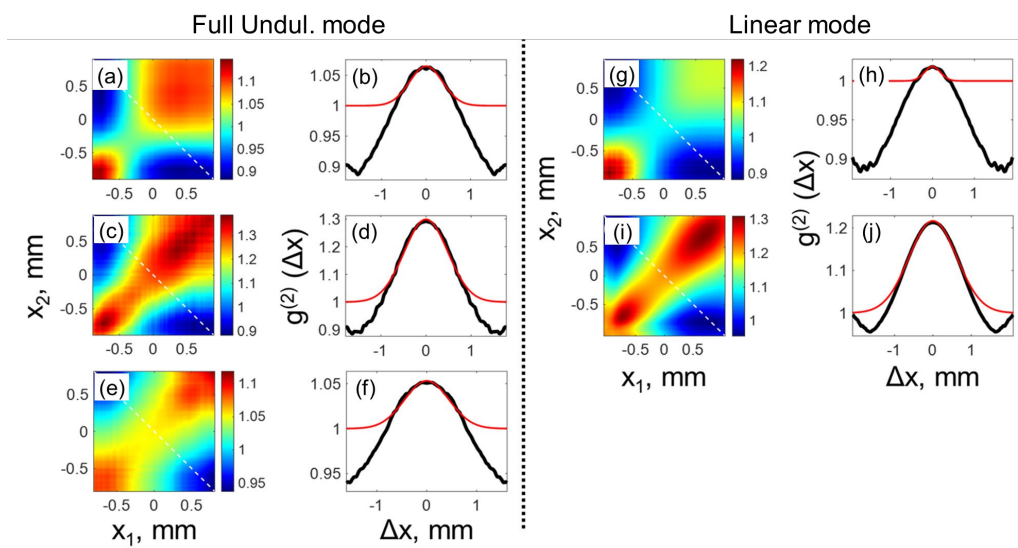

**Figure S13** Spatial $g^{(2)}$-function analysis along the horizontal direction and one-dimensional anti-diagonal profile (along white dash line) for operation of the PAL XFEL source at the 120 pC bunch charge. Left side: (a,b) SASE mode, (c,d) SASE monochromatic mode, and (e,f) self-seeding mode of operation. Right side, linear mode of operation: (g,h) SASE mode, (i,j) SASE monochromatic mode. The red lines are the results of Gaussian fitting.

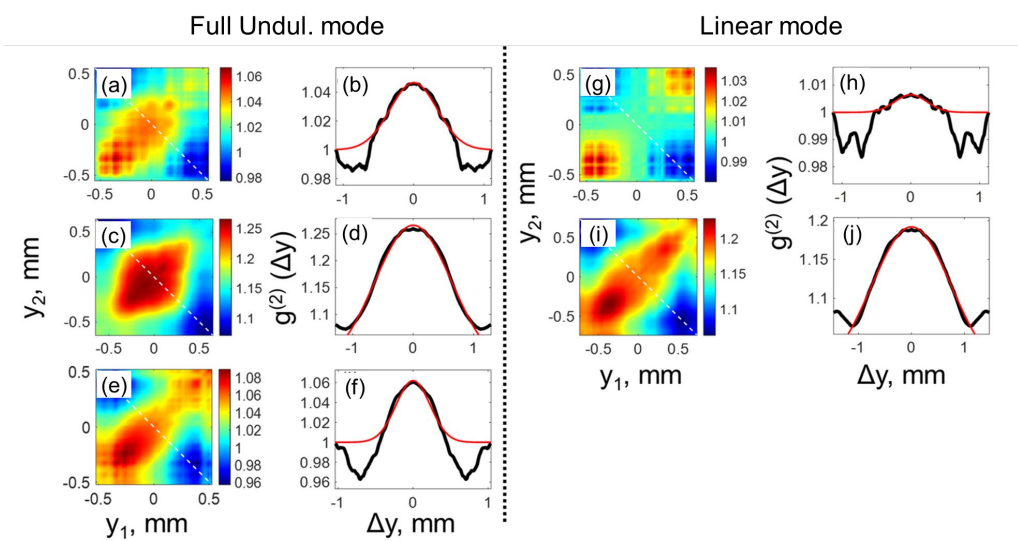

**Figure S14** Spatial $g^{(2)}$-function analysis along the vertical direction and one-dimensional anti-diagonal profile (along white dash line) for operation of the PAL XFEL source at the 120 pC bunch charge. Left side: (a,b) SASE mode, (c,d) SASE monochromatic mode, and (e,f) self-seeding mode of operation. Right side, linear mode of operation: (g,h) SASE mode, (i,j) SASE monochromatic mode. The red lines are the results of Gaussian fitting.

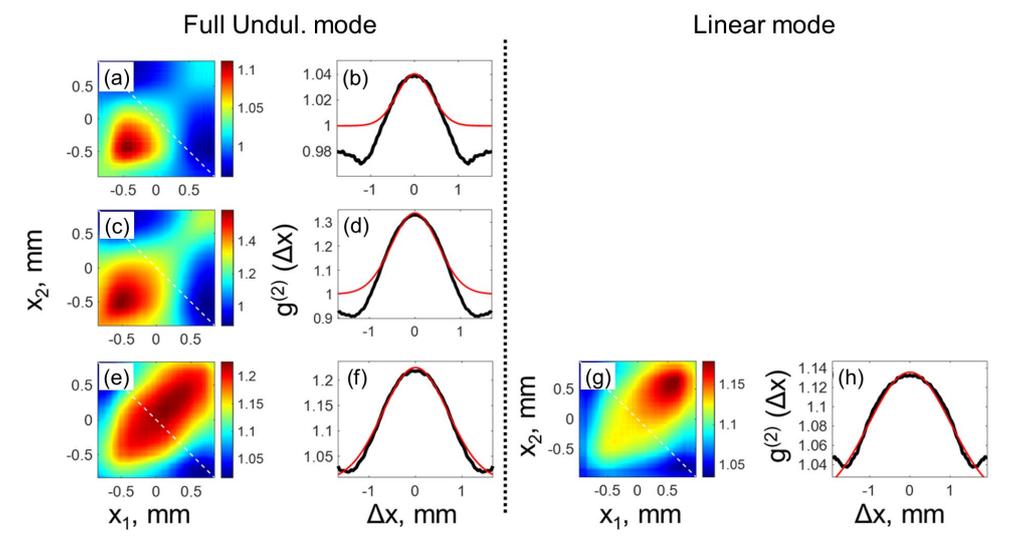

**Figure S15** Spatial $g^{(2)}$-function analysis along the horizontal direction and one-dimensional anti-diagonal profile (along white dash line) for operation of the PAL XFEL source at the 200 pC bunch charge. Left side: (a,b) SASE mode, (c,d) SASE monochromatic mode, and (e,f) self-seeding mode of operation. Right side, linear mode of operation: (g,h) self-seeding mode of operation. The red lines are the results of Gaussian fitting.

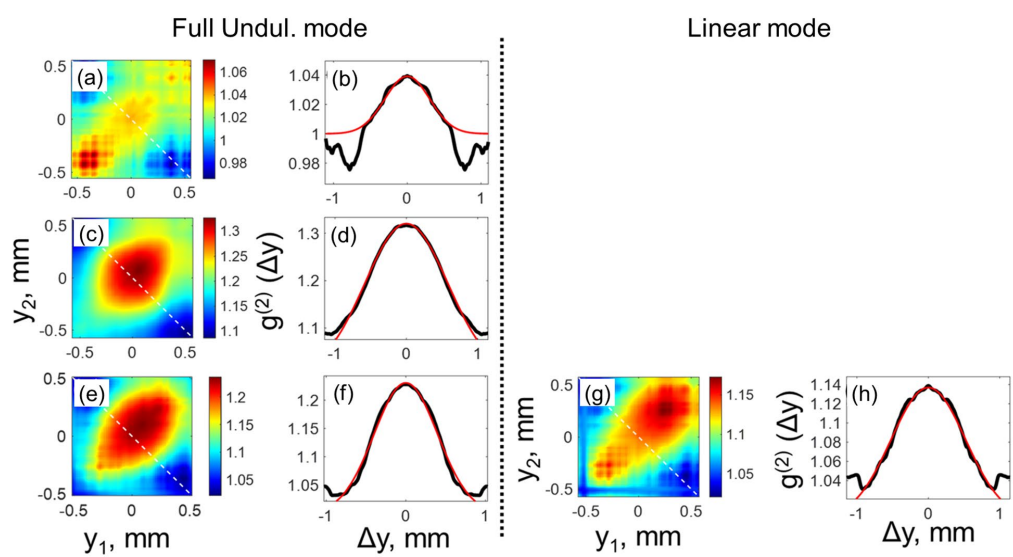

**Figure S16** Spatial $g^{(2)}$-function analysis along the vertical direction and one-dimensional anti-diagonal profile (along white dash line) for operation of the PAL XFEL source at the 200 pC bunch charge. Left side: (a,b) SASE mode, (c,d) SASE monochromatic mode, and (e,f) self-seeding mode of operation. Right side, linear mode of operation: (g,h) self-seeding mode of operation. The red lines are the results of Gaussian fitting.

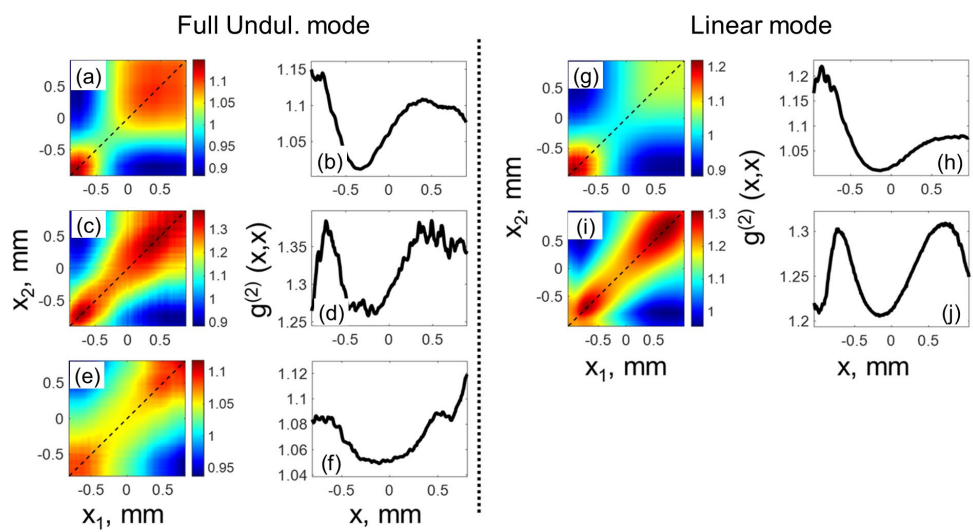

**Figure S17** Spatial $g^{(2)}$-function analysis along the horizontal direction and one-dimensional diagonal profile (along black dash line) for operation of the PAL XFEL source at the 120 pC bunch charge. Left side: (a,b) SASE mode, (c,d) SASE monochromatic mode, and (e,f) self-seeding mode of operation. Right side, linear mode of operation: (g,h) SASE mode, (i,j) SASE monochromatic mode.

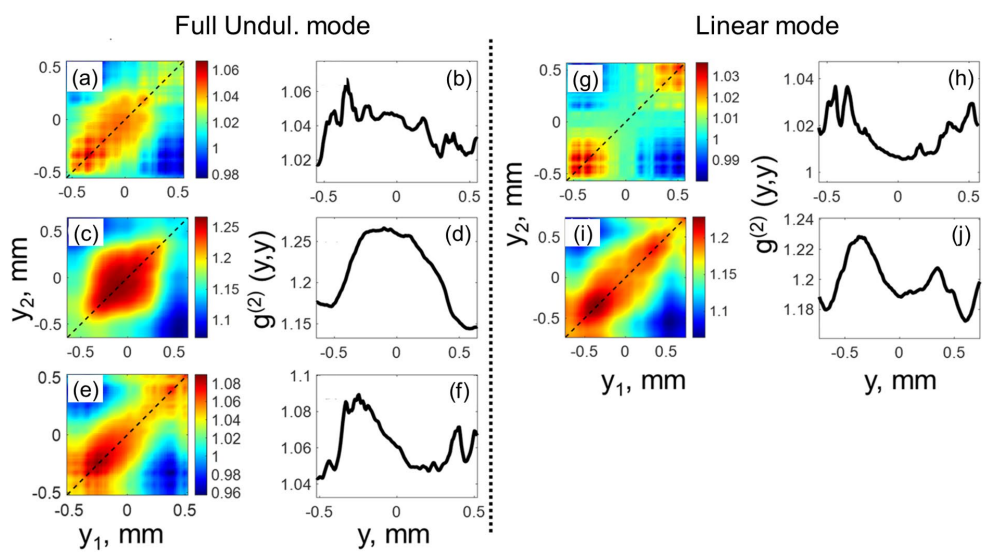

**Figure S18** Spatial $g^{(2)}$-function analysis along the vertical direction and one-dimensional diagonal profile (along black dash line) for operation of the PAL XFEL source at the 120 pC bunch charge. Left side: (a,b) SASE mode, (c,d) SASE monochromatic mode, and (e,f) self-seeding mode of operation. Right side, linear mode of operation: (g,h) SASE mode, (i,j) SASE monochromatic mode.

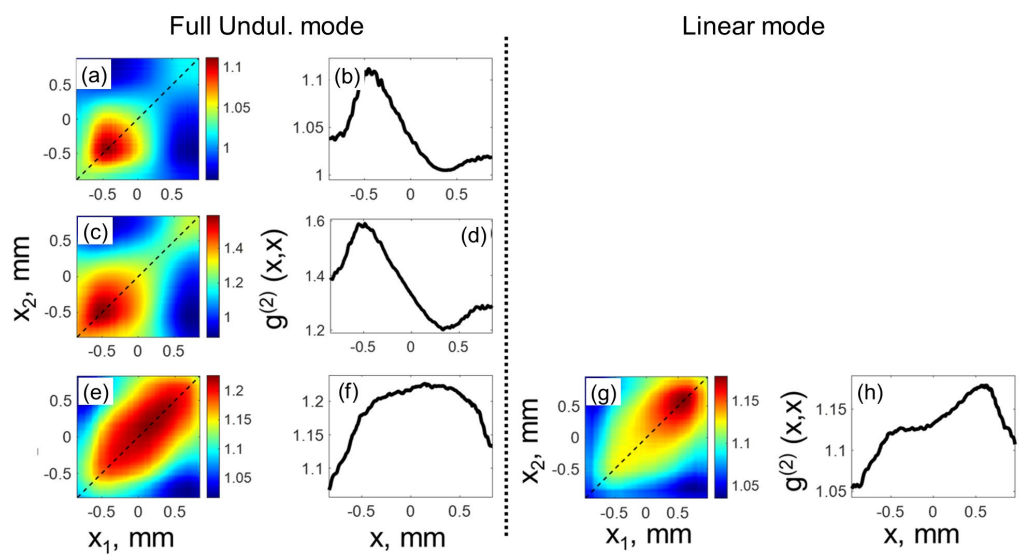

**Figure S19** Spatial $g^{(2)}$-function analysis along the horizontal direction and one-dimensional diagonal profile (along black dash line) for operation of the PAL XFEL source at the 200 pC bunch charge. Left side: (a,b) SASE mode, (c,d) SASE monochromatic mode, and (e,f) self-seeding mode of operation. Right side, linear mode of operation: (g,h) self-seeding mode of operation.

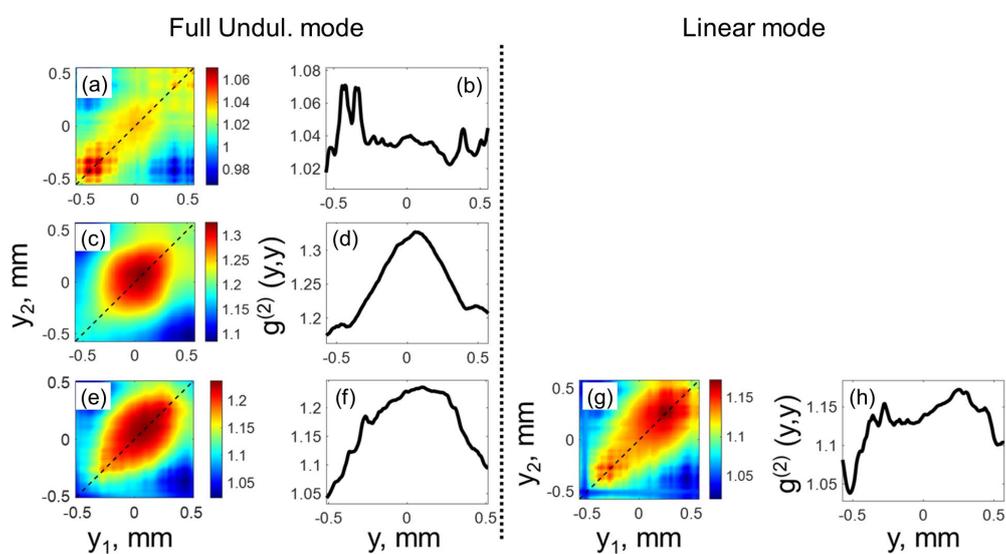

**Figure S20** Spatial $g^{(2)}$-function analysis along the vertical direction and one-dimensional diagonal profile (along black dash line) for operation of the PAL XFEL source at the 200 pC bunch charge. Left side: (a,b) SASE mode, (c,d) SASE monochromatic mode, and (e,f) self-seeding mode of operation. Right side, linear mode of operation: (g,h) self-seeding mode of operation.